\documentclass[aps,pre,floatfix,reprint,twocolumn,superscriptaddress,showpacs,titlepage,amssymb,amsmath]{revtex4-2}

\usepackage{graphicx, times, epsfig, color, float, mathtools}
\usepackage{subfigure}
\usepackage{xcolor}
\usepackage[colorlinks = true, linkcolor = blue, urlcolor  = blue, citecolor = blue, anchorcolor = blue]{hyperref} 

\usepackage{adjustbox}
\usepackage{tikz}
\usetikzlibrary{matrix,shapes,arrows,positioning,chains}
\tikzstyle{box} = [rectangle, rounded corners, minimum width=3cm, minimum height=1cm,text centered, draw=black, fill=green!10]
\tikzstyle{arrow} = [thick,->]

\def\be{\begin{equation}}
\def\ee{\end{equation}}
\def\bea{\begin{eqnarray}}
\def\eea{\end{eqnarray}}
\def\etal{{\it et al.}}
\def\degree{$^{\circ}$}
\def\atp{at.\,\%}
\def\ath{at.\,\%\,H}

\def \asihx {{\it a}-Si$_{\text{1-x}}$H$_{\text{x}}$} 
\def \asih {{\it a}-Si:H}
\def \asi {{\it a}-Si}

\begin{document} 
\title{
Ab initio study of the structure and properties of amorphous silicon 
hydride from accelerated molecular dynamics simulations
}

\author{Raymond Atta-Fynn} 
\email{Corresponding author:\,attafynn@lanl.gov}
\affiliation{Department of Physics, University of Texas, Arlington, Texas 76019}
\thanks{Present address: {\it Los Alamos National Laboratory, Los Alamos, NM, 87545, USA}}

\author{Somilkumar J. Rathi} 
\email{so954994@ucf.edu}
\affiliation{Department of Materials Science and Engineering, University of Central Florida, Orlando, Florida 32816}

\author{Harsh Arya}
\email{harsh.arya@mavs.uta.edu}
\affiliation{Department of Physics, University of Texas, Arlington, Texas 76019}

\author{Parthapratim Biswas}
\email{partha.biswas@usm.edu}
\affiliation{Department of Physics and Astronomy, University of Southern 
Mississippi, Hattiesburg, Mississippi 39406}

\begin{abstract}
This paper presents a large-scale {\it ab initio} simulation 
study of amorphous silicon hydride ({\asihx}) with an 
emphasis on the structure and properties of the material 
across a range of hydrogen concentration 
by combining accelerated molecular dynamics (MD) simulations 
with first-principles density-functional calculations. 
The accelerated MD scheme relied on classical metadynamics, 
which enabled the development of 2600+ high-quality 
structural models of {\asihx}, with system sizes ranging 
from 150 to 6,000 atoms and hydrogen concentrations 
vary from 6 to 20 {\atp}.  The resulting amorphous 
networks were found to be completely free from any coordination 
defects and that they all exhibited a pristine band-gap in their 
electronic spectrum. The microstructural properties of 
hydrogen distributions were examined with great emphasis 
on the presence of isolated and clustered environments 
of hydrogen atoms. The results were compared with a suite of 
experimental data obtained from x-ray diffraction, infrared 
spectroscopy, spectroscopic ellipsometry and nuclear 
magnetic resonance studies. 
\end{abstract}

\keywords{amorphous silicon, structural database, 
molecular dynamics simulation}

\maketitle 
\section{Introduction}
Hydrogenated amorphous silicon ({\it a}-Si:H) is an 
important semiconducting material with a range of 
applications to material devices, especially 
in photovoltaics~\cite{HIT1, pv-module,HIT}. 
However, the photovoltaic properties of 
{\asih}-based solar cells are adversely affected 
by the so-called Staebler-Wronski effect~\cite{SW1,SW2}. 
The problem involves a gradual degradation of 
the cell upon prolonged exposure to sunlight 
and the consequent loss of its photovoltaic 
efficiency with time. 
Although the exact cause of this degradation is 
largely unknown and still a debating issue, it is widely 
accepted  that the light-induced structural changes 
associated with the breaking of certain 
Si--H bonds and the subsequent motion of H 
atoms in the network play a central role in 
the degradation process. 
The elucidation of such effects, via computer modeling 
and theoretical analyses of simulation results, 
requires large-scale, realistic, device-quality models of 
{\it a}-Si:H, which can produce not only the correct 
structural~\cite{ECMR07}, electronic~\cite{Meta2016} and 
microstructural properties~\cite{Biswas2017,Biswas2011}
of hydrogen distributions in {\asih}, but also 
the dynamics of H atoms in the network~\cite{Biswas2020,Biswas2021}
 
In the last decade, several smart simulation techniques 
were developed that can outperform the conventional 
methods~\cite{Min,Holender,Tuttle,buda,Jarolimek,drabold1,klein},
based on Monte Carlo and molecular-dynamics simulations,
to produce high-quality models of disordered materials.
These techniques can be collectively described as data-driven 
methods, which are thematically related to each other 
by their ability to employ either training data sets 
obtained from prior calculations to construct 
a knowledge-based potential or structural information 
derived from experiments to develop data-assisted 
simulation strategies. Examples of these techniques 
include, but are not limited to,  machine-learning~(ML) 
approaches~\cite{Deringer2018} 
and a number of experimental information based 
pure~\cite{RMC04,RMC96,CMC19} and hybrid 
approaches~\cite{ECMR05,ECMR07,FEAR-SR,FEAR-PRB,INDIA}, 
developed in the context of modeling disordered solids. 
The so-called ML approaches rely on the availability of 
accurate training data, which are generally, but not 
necessarily, obtained from a small number of first-principles 
calculations on limited system sizes. 
As such, the success of ML methods largely depends on the 
quality and quantity of training data, as well as 
their ability to represent the configuration 
space of very many structural solutions associated with 
disordered solids.
By contrast, the effectiveness of data-assisted hybrid 
approaches hinges on the information content of experimental 
data -- from diffraction and other spectroscopic 
measurements -- and the feasibility of employing scalar/vector 
information in association with approximate force fields or total-energy 
functionals to form an augmented solution space. The approach 
then seeks suitable structural solutions in the augmented 
space in order for the solutions to be satisfied by theory 
and experiments simultaneously. 
However,  the use of {\it ab initio} force fields (in ML methods) 
or the lack of suitable classical/semi-classical total-energy 
functionals and adequate experimental data (for hybrid approaches) can 
restrict these methods from applying to large multicomponent 
disordered systems, with few exceptions. 

In view of the preceding observation, the development 
of large realistic models of amorphous solids continues to 
pose a major problem. For {\asih}, the key 
difficulties include: 1) the generation of 
high-quality atomistic configurations of {\asi}; 
2) the creation of appropriate types and numbers 
of silicon-hydrogen bonding configurations, which
depend on the concentration of hydrogen in the 
network; and 3) the lack of suitable classical or 
semi-classical potentials to accurately describe 
the interaction between Si and H atoms. The last 
issue is particularly problematic for large systems 
as it is necessary to hydrogenate {\asi} networks to produce a 
correct hydrogen distribution. 
Although a number of ad hoc approaches~\cite{srepb} 
have been developed to hydrogenate {\asi} networks with 
coordination defects, none of these approaches 
so far can address these issues satisfactorily, 
especially the creation of a variety of defective 
structures of {\asi} to generate the correct concentration 
dependent hydrogen microstructure. These issues call 
for the development of an efficient 
method that can yield {\asih} models 
with all possible silicon-hydrogen bonding 
configurations across a wide range of hydrogen concentration. 

The goal of this study is to develop 
a structural database of {\asih}, consisting of thousands 
of amorphous configurations, with system sizes vary 
from 150 atoms to 6000 atoms and hydrogen concentrations 
in the range of 6--20 {\atp}.  The approach adopted here 
is based on a combination of conventional molecular-dynamics 
simulations and their accelerated counterpart, known 
as metadynamics.  We have shown that the approach can successfully
address the issues mentioned earlier for fairly 
large systems using {\it ab initio} total-energy minimization. 
Here, we have systematically applied this approach to 
generate a large ensemble of {\asih} configurations with 
varying system sizes and hydrogen concentrations. The efficiency 
of our approach emanates from its ability to control 
atomic coordination of Si, and thereby producing a variety 
of Si--H bonding configurations, the properties of which are found 
to be in good agreement with experimental data from 
nuclear magnetic resonances (NMR)~\cite{Taylor,nmr1,nmr2,Baum}, 
infrared (IR) spectroscopy~\cite{Chabal1984,Scharff}, 
spectroscopic ellipsometry~\cite{Kageyama}, and 
inelastic neutron scattering measurements~\cite{Kamita}. 

The remainder of this paper is organized as follows.
In Sec.~\ref{comp}, we describe the simulation method 
in detail, which comprises the generation of an ensemble 
of defective {\asi} structures via metadynamics, followed 
by hydrogen passivation of defects and the total-energy 
minimization of the resulting {\it a}-Si:H structures 
using density functional theory (DFT). 
Section~\ref{results} discusses the structural and 
microstructural properties of {\asih} models and 
the distribution of H atoms in the networks in 
relation to the shape of the nuclear magnetic resonance 
(NMR) line spectra of {\asih}.  The phonon and 
electro-optical properties of {\asih} are also 
discussed and compared with experimental data.  
Section~\ref{website} provides information on the 
structural database that resulted from our study. 
It incorporates detailed structural, electronic and 
additional information on a few thousand atomic 
configurations of {\asih}. This is followed by 
conclusions in Sec.~\ref{conclusion}.

\section{Computational Methodology} \label{comp}

\begin{figure}[tbp!]
\includegraphics[height=4.0 in, width=2.5 in]{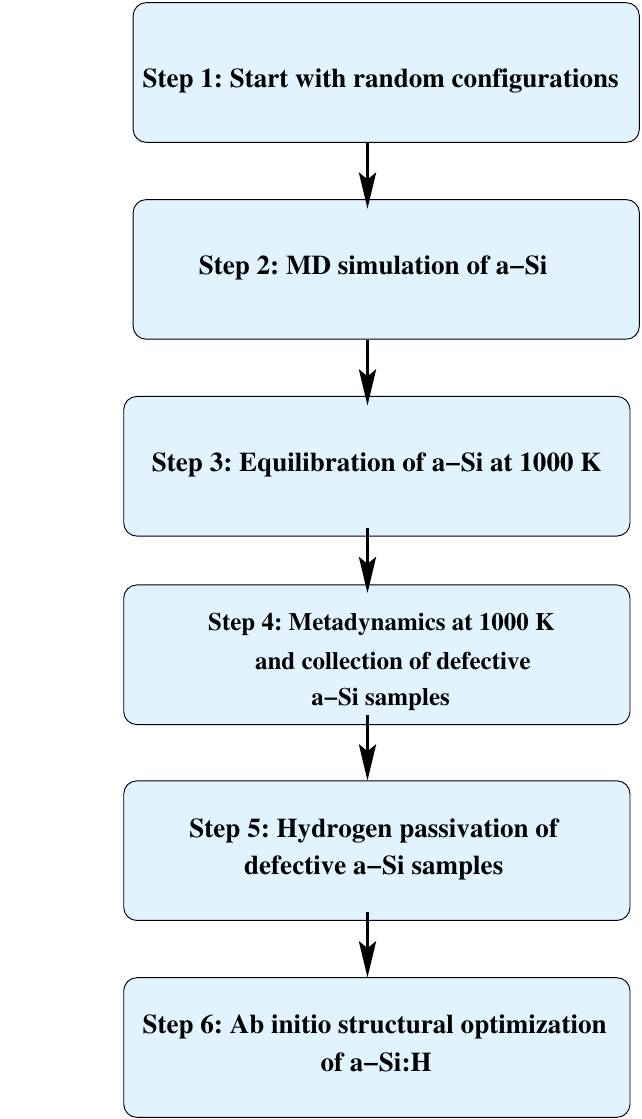}
\caption{\label{fchart}
A flowchart depicting the key steps of the method 
employed in this study to generate {\asih} models 
using conventional and accelerated MD simulations, 
followed by the passivation of defects with hydrogen. 
}
\end{figure}

In this study, we employed a combination 
of conventional molecular-dynamics (MD) simulations and 
an accelerated MD technique to 
generate an array of structural 
configurations of {\asi} spanning a wide range of concentrations 
of coordination defects.  The configurations were 
then passivated with H atoms to produce models of {\asih}. 
A flowchart of the entire 
procedure is depicted in Fig.~\ref{fchart}, which lays 
out the key steps of the method.  Below, we give a description 
of the accelerated MD technique, which is followed 
by a brief discussion of classical MD simulations 
of {\asi} and the subsequent development of {\asih} 
models via hydrogen passivation. 

\subsection{Accelerated Molecular Dynamics}

The primary goal of the accelerated MD technique is 
to produce representative structural configurations 
of {\asi} with undercoordinated atoms or defects, which 
play a key role in producing {\asih} models. The ability to produce a 
structural configuration with a given number and type of 
undercoordinated atoms provides a means to generate various 
silicon-hydrogen bonding configurations that have been 
experimentally observed in {\asih} for a wide range of 
H concentration. The accelerated MD approach used in this study 
is based on the metadynamics method~\cite{Laio}.
Metadynamics is a technique that provides a fast sampling 
of the free-energy surface associated with an event or 
mechanism, which is difficult to access via conventional 
MD. In the present context, we are primarily interested 
in the low-energy structural configurations of {\asi} that 
correspond to a given number (and type) of undercoordinated 
Si atoms. 

The principles of metadynamics can be summarized 
as follows:\\ (i) First, a small set, 
${\bf s}$, of differentiable functions, $s_\ell$, 
of the system coordinates ${\bf \mathbf{R}}$, is 
chosen so that they can describe the mechanism 
of interest through 
\be
{\bf s}=\{s_\ell({\bf \mathbf{R}}):\ell=1,2,\dots,d\}. 
\ee
The functions $\{s_\ell({\bf \mathbf{R}})\}$ are known 
as collective variables. The dimensionality, $d$, of 
the collective variables is typically in the range 
$1\leq d \leq 3$.\\ 
(ii) 
Second, a history (i.e., time-dependent) potential, 
$\mathcal{V}({\bf s}, t)$, is periodically 
added to the Hamiltonian of the system during 
simulation, where $t$ denotes time. 
The periodic addition of $\mathcal{V}({\bf s},t)$ effectively discourages 
the system from visiting regions of the free-energy 
surface that have already been visited, and thus 
drives the system to explore new regions of the free-energy 
surface. As the system evolves, $\mathcal{V}$ is 
periodically updated by adding a $d$-dimensional 
$\delta$ function centered on the value of ${\bf s}$. 
To put it more succinctly, suppose that at a given time 
$t$ during metadynamics simulation, $\mathcal{V}$ is updated 
$n$ times at regular time periods $\tau,\,2\tau,\,3\tau,\dots,n\tau$ 
($t\geq n\tau$). If the set of collective variables 
at these time periods is specified by $\{ {\bf s}_\tau,\,
{\bf s}_{2\tau},\,{\bf s}_{3\tau},\dots,{\bf s}_{n\tau}\}$, 
then the history potential, $\mathcal{V}({\bf s},t)$, 
for an arbitrary set of collective variables ${\bf s}$ 
at time $t$, is defined as 
\be
\mathcal{V}({\bf s},t)=\sum_{k=1}^{n}\delta({\bf s}-{\bf s}_{k\tau})=\sum_{k=1}^{n}\prod_{\ell=1}^{d}\delta(s_{\ell}-s_{\ell_{k\tau}}); 
\label{metav}
\ee
(iii) Third, the basic idea of metadynamics is that, in the 
presence of bias and true potentials, after a sufficiently 
long simulation time period, the free energy 
$\mathcal{F}({\bf s})$ of the system spanned by the set 
of collective variables ${\bf s}$, can be recovered from 
the history potential $\mathcal{V}({\bf s},t)$
\be
\mathcal{F}({\bf s})=-\lim_{t\to\infty}\mathcal{V}({\bf s},t). 
\ee

It is customary to approximate the delta function, $\delta(x-x_0)$, 
in Eq.~(\ref{metav}) by a narrow Gaussian function $g(x)$ of height 
$A$ (with the dimension of energy) and width $\sigma$ (with the 
dimension of the collective variable), which is 
centered on $x_0$
\be
g(x;x_0,\sigma)=A\exp\left[-\frac{(x-x_0)^2}{2\sigma^2}\right]. 
\label{gauss}
\ee
For large system sizes, where thousands of Gaussian functions 
are required to efficiently sample the energy surface on 
the fly, Lucy's function~\cite{Lucy_1977} often serves as an accurate but 
computationally cheaper alternative to Gaussian functions. 
Writing $y = x - x_0$, Lucy's function, $\mathcal{L}$, is
defined as 
\be
 \mathcal{L}(y;w)=
\begin{dcases}
    \!A\!\left[\!1\!+\!\frac{3y}{w}\!\right]
\!\left[\!1-\frac{y}{w}\!\right]^3, 
 &\text{for} \: \lvert y \rvert \: \le{w}\\
    0,              & \text{otherwise}
\end{dcases}
\label{Lucy}
\ee
where $A$ is the height of the function~\cite{normalization},
$x_0$ is the center, and $w>0$ is the support 
radius of the function. It can 
be shown that the Gaussian function, $g(x;x_0,\sigma)$, and 
Lucy's function, $\mathcal{L}(x;x_0,w)$, in Eqs.~(\ref{gauss}) 
and (\ref{Lucy}), respectively, are computationally 
equivalent~\cite{lucy} for $w=3.052444\sigma$ in the 
region $\lvert x - x_0\rvert \le w$.  
In other words, a Gaussian function of height $A$ and width 
$\sigma$, which is centered on $x_0$, is equivalent to 
Lucy's function of height $A$ and width $w$, centered 
on $x_0$ in the interval $\lvert x-x_0\rvert\!\le{w}$. 
In this work, a $\delta$ function is approximated 
by Lucy's function. 
Thus, the history potential in Eq.~(\ref{metav}) 
becomes 
\be
\mathcal{V}({\bf s},t)=\sum_{k=1}^{n}\prod_{\ell=1}^{d}\mathcal{L}(s_{\ell}; s_{\ell_{k\tau}},w). 
\label{metav2}
\ee
In Sec.~IIB, we briefly describe how this accelerated MD 
technique can be effectively combined with conventional MD 
simulations to obtain a set of desired {\asi} configurations 
for hydrogen passivation. 

\subsection{Sampling energy landscape of {\asi}, 
hydrogen passivation, and {\it ab initio} thermalization and 
relaxation}

We employed classical MD simulations to produce 
{\asi} models. Starting with a set of independent 
random initial configurations within a cubic 
supercell, an array of amorphous silicon configurations 
with a density of 2.25 g\,cm$^{-3}$ were 
generated.  The size of the models ranges from 
150 to 6,000 atoms and the concentration of atomic 
hydrogen lies between 6 to 20 at.~\%. The choice 
of model sizes and hydrogen concentrations is 
motivated by our desire to create an accurate 
structural database of {\asih} for future use.

Since amorphous silicon is not a glass, direct 
applications of MD simulations using the so-called 
melt-quench approach tend to produce too many 
coordination defects in the networks. To avoid this 
problem, we adopted a specially developed MD 
protocol by Atta-Fynn and Biswas~\cite{raf_asi},
which employed the modified Stillinger-Weber 
potentials (SW)~\cite{SW-silicon,SW-Vink} to generate 
high-quality structural configurations of {\asi}. 
The resulting {\asi} 
configurations were subjected to annealing at 
1000 K for a period of 100 ps, and the annealed 
models were used as the starting configurations 
of several independent metadynamics runs at 
1000~K~\cite{MTD}. To generate different types and numbers 
of undercoordinated atomic configurations of 
{\asi}, for the purpose of hydrogen passivation, 
we define a collective variable, $s_i(t)$, which 
plays the role of the atomic coordination number 
at site $i$
\be 
s_i = \sum_{\substack{j=1 \\(j \ne i)}}^N\mathcal{W}(r_{ij}).
\ee 
Here, $r_{ij}=|r_i-r_j|$, where $r_i$ and $r_j$ are the atomic 
positions at sites $i$ and $j$, respectively, and 
$\mathcal{W}(r_{ij})$ is a polynomial 
function that decays smoothly from 1 to 0 for 
increasing values of $r_{ij}$ from 0 to $r_c$. 
The function $\mathcal{W}(r_{ij})$ is given by 
 
\[\mathcal{W}(r_{ij})=
\begin{dcases}
1, &\text{if } r_{ij}\le r_0 \\
2\left[t(r_{ij})\right]^3 -3\left[t(r_{ij})\right]^2 + 1, & \text{if } r_0 < r_{ij} < r_c \\
    0,              & r_{ij}\ge r_c
\end{dcases}
\] 
\noindent with
\be
t(r_{ij})=\frac{r_{ij}-r_0}{r_c-r_0}.
\ee
This definition allows us to use a smooth continuous 
variable as the coordination number of an atom 
(without introducing a sharp cutoff value for 
atomic coordination) during metadynamics runs 
when the distance between neighboring atoms 
fluctuate considerably. The constants $r_0$ and $r_c$ 
correspond to the lower and upper boundaries 
of the zero region after the first shell in 
the pair-correlation function of {\asi}.  In 
this study, we employed $r_0=2.7$ {\AA} and 
$r_c=3.0$ {\AA}. The average coordination 
number, $s(t)$, of a network at time $t$ is 
obtained by averaging over all sites 
\be 
s(t) = \frac{1}{N}\sum_{i=1}^N s_i(t).  
\label{avg} 
\ee

\begin{figure}[tbp!]
\includegraphics[clip,width=0.9\linewidth]{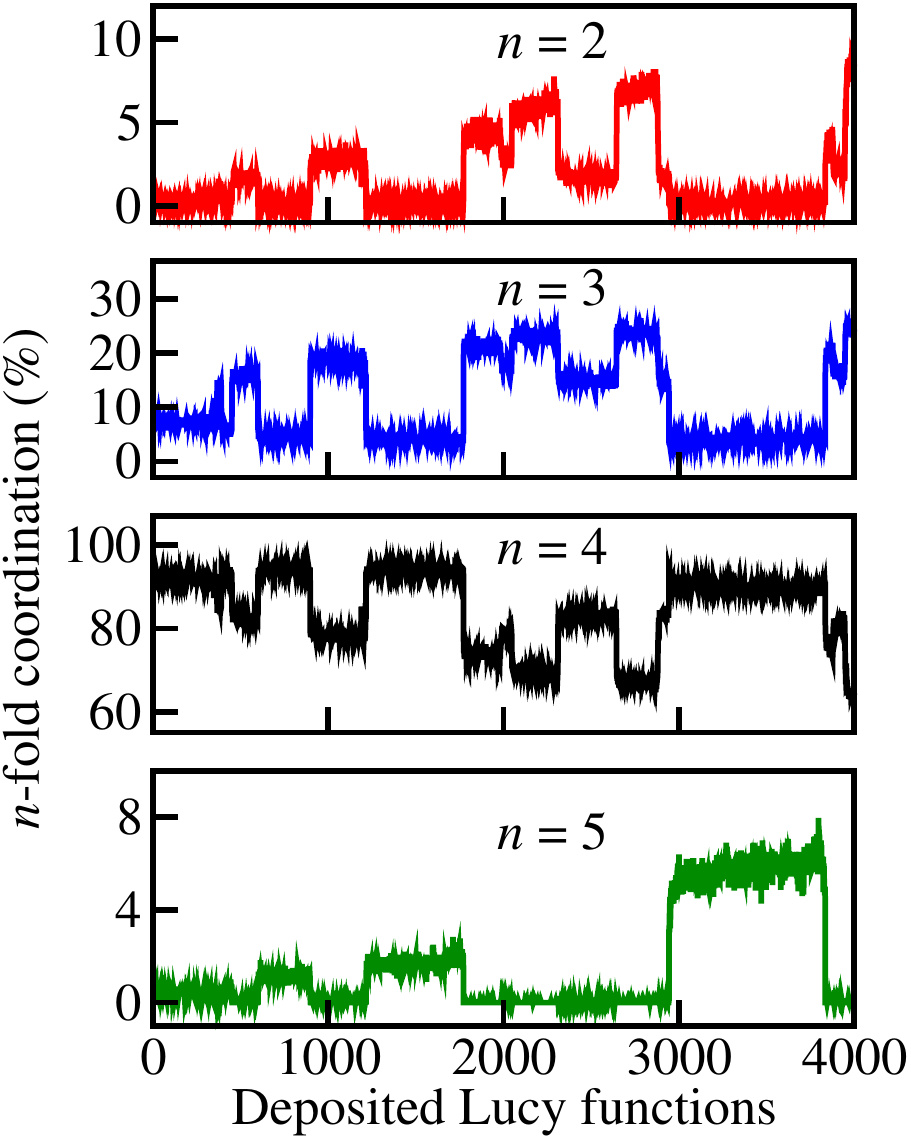}
\caption{\label{nfig1}
The evolution of the number of $n$-fold-coordinated 
atoms (in percent) during metadynamics simulations 
using the modified Stillinger-Weber 
potential at 1000 K in a 1000-atom model of {\asi} 
with the number of Lucy's functions, deposited at the rate 
of one function per 5 ps.  
}
\end{figure}

The metadynamics simulations were performed by adding a Lucy's 
function of height $A$=0.1 eV and width $w$=0.305 
to the bias potential $\mathcal{V}$ at a periodic 
interval of $\tau=5$~ps in order to drive the 
system through different regions of the free-energy 
landscape associated with the collective variable 
$s$. As the system 
propagates under the influence of $\mathcal{V}$ on the 
Stillinger-Weber potential-energy surface, one can track the 
evolution of the average coordination number, $s(t)$, 
as a function of time. This is shown in 
Fig.~\ref{nfig1}, where the variation of the number 
of $n$-fold-coordinated atoms (in percent) is 
plotted against the number of Lucy's functions 
deposited at time $t$.  
Thus, atomic configurations with a given number 
and type of defects, such 
as one-fold-, two-fold-, and three-fold-coordinated 
atoms, can be readily harvested along 
metadynamic trajectories.  The defective
configurations of {\asi} are so chosen that 
the concentration of H atoms lies within the 
range from 6 to 20 {\atp} after the defects 
are hydrogenated. Once the defective configurations 
are in place, the passivation of defects proceeds 
by inserting H atoms in the vicinity of the 
defective sites such that the resulting local 
structure is approximately tetrahedral. The 
supercell volume is then adjusted to ensure that 
the mass density of the resultant {\asih} models 
agrees with the experimental density of 
{\asih}~\cite{Smets}.

Although the {\asih} models obtained from the procedure 
described above yield a realistic distribution of hydrogen 
atoms, the inclusion of H atoms near defective sites 
may induce some strain in amorphous networks. To 
minimize the effect of this strain on the hydrogen 
microstructure of {\asih}, the resulting structures were 
thermalized at 300~K for 3--5 ps, followed by structural 
relaxation, using the {\it ab initio} DFT
code {\sc Siesta}~\cite{siesta}. The thermalizations and 
structural relaxations of the models were conducted 
using both the self-consistent-field (SCF) and non-SCF 
approximations. For systems consisting of more than 
3000 Si atoms, we employed the non-SCF Harris functional 
approach~\cite{harris} for solving the Kohn-Sham 
equation. It has been shown elsewhere~\cite{Ray-PRB} 
that for {\asi} the results from the Harris approach 
are on a par with those from full SCF calculations. 
The Ceperley-Alder formulation~\cite{Ceperley1980} 
of the local density approximation (LDA) was used to 
account for the exchange-correlation energy of 
the system, and the norm-conserving pseudopotentials 
in the Troullier-Martins form were used to describe 
the electron-ion interactions~\cite{tm}. 
Double-zeta basis functions were used for systems 
up to 3,000 atoms, whereas systems with more than 
3,000 atoms are treated using the single-zeta basis 
functions. Only the $\Gamma$ point, $\vec{k}$={\bf 0}, 
was used to perform the Brillouin-zone integration. 
The relaxation procedure continued until the total 
force on each atom was found to be less than 10 
meV/{\AA}. 

\begin{figure}[tbp!]
\includegraphics[clip,width=0.9\linewidth]{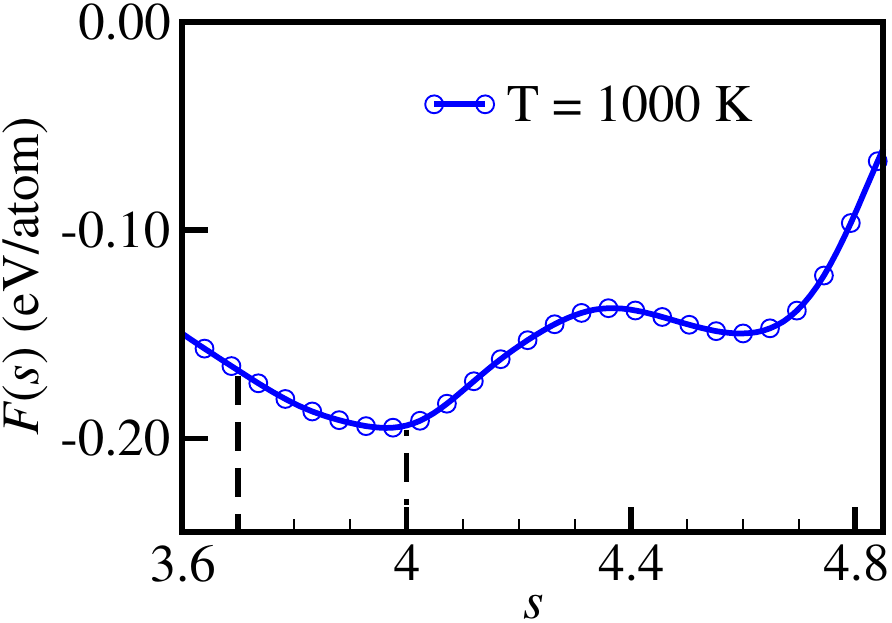}
\caption{\label{nfig3}
The variation of the free energy (blue circles) 
per atom, $F(s)$, with the average atomic coordination 
number, $s$, for a 1000-atom {\asi} configuration 
during metadynamics simulation at 1000~K.  The two 
vertical lines (black) represent the coordination 
region from where the defective configurations were 
collected for hydrogenation.  The minimum of 
$F(s)$ corresponds to the value of $s=3.94$.
}
\end{figure}

\section{Results and Discussion}\label{results}

This section provides a complete characterization of 
{\asihx} models with hydrogen concentrations in the 
range from 6 to 20 at.~\%. 
Since it is not possible to include a comprehensive 
discussion of models for all hydrogen concentrations
in the limited space, we have restricted our discussion 
to a set of models with representative low (6--9 {\atp}), 
intermediate (10--13 {\atp}) and high (18--20 {\atp}) 
concentration of hydrogen in presenting the results. A similar 
observation applies to model sizes as well. The models 
with the low concentration of H are reflective of 
device-grade samples of {\asih}, whereas the models 
with the high concentration of H provide rich microstructural 
information of hydrogen distribution in amorphous silicon 
networks involving silicon-hydrogen bonding configurations 
and isolated/clustered environments of H atoms, as 
observed in infrared (IR)~\cite{Chabal1984,RS}, 
nuclear magnetic resonance (NMR)~\cite{nmr1,Baum} 
and hydrogen-effusion measurements~\cite{Beyer2003}. 
In some cases, our choice of H concentration was simply 
dictated by experimental data with given H concentrations. 
The quality of the models are studied by analyzing the structural, 
electronic, vibrational and microstructural 
properties of hydrogen distributions in the 
networks.  The results are compared with available 
experimental data.  

\subsection{Structural Properties of {\asih}}
Figure~\ref{nfig3} depicts the evolution of the free energy, 
$F(s)$, with the average coordination number, $s$.  A close 
examination of the plot reveals that the minimum of $F(s)$ 
corresponds to $s=3.94$ at 1000~K, which is 
slightly less than the ideal tetrahedral value of 4. This 
implies that at high temperatures some of the atoms in 
{\asi} models can be undercoordinated. The value of $s$ 
is close to the experimental value of 3.88, obtained from x-ray 
diffraction measurements on annealed samples of {\asi} at 
873~K by Laaziri {\etal}~\cite{laaz}
It may be noted that this result is particularly 
true for bulk samples, which are characterized by 
the presence of a few vacancy-type defects in the networks. For 
finite-size computer-generated models, a value of 3.88 at 
the room temperature (of 293~K) generally indicates 
the presence of too many {\em isolated} dangling bonds, 
which can produce unwanted defect states in the 
electronic density of states and lead to a pseudo-gap or 
gapless spectrum~\cite{note1}. In the present 
case, an average coordination value of 3.94 in 
metadynamical models is indicative of continuous
bond formation and dissociation~\cite{Biswas2021} 
in the network at 1000~K as the system evolves 
on the high temperature free-energy surface.

\begin{figure}[tbp!]
\includegraphics[clip,width=0.9\linewidth]{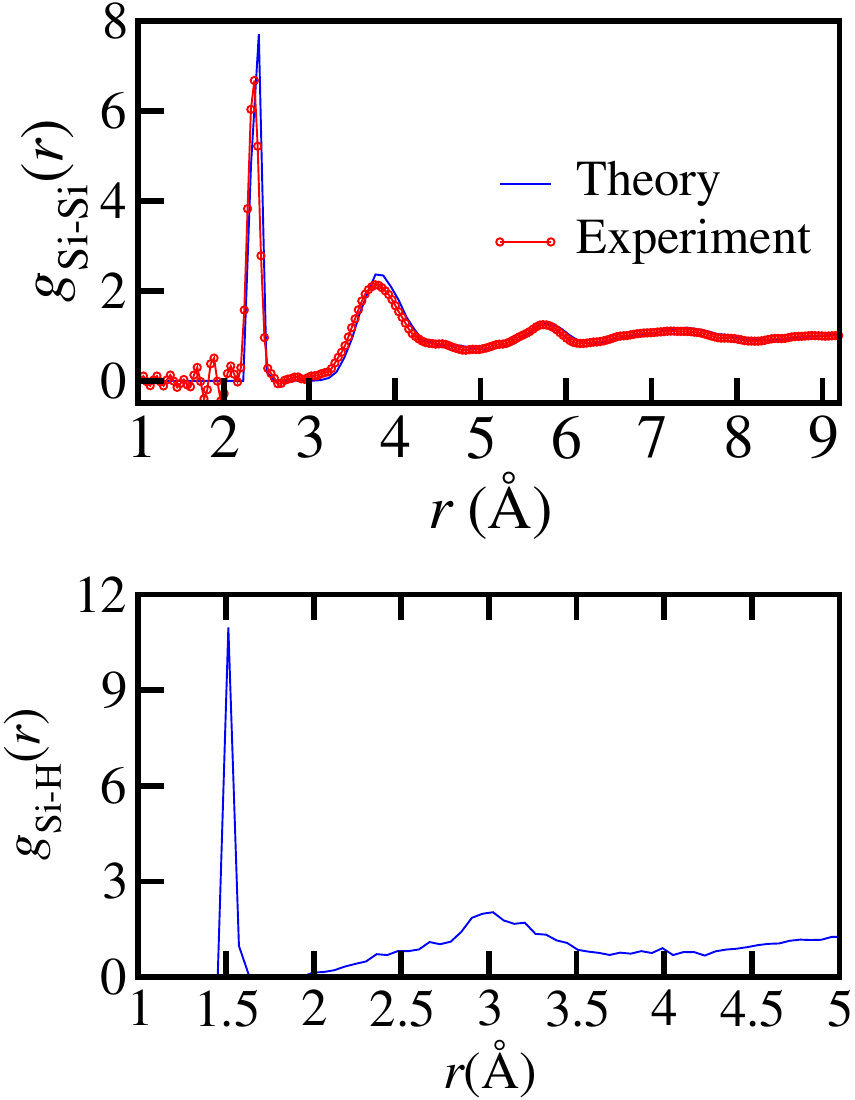}
\caption{\label{nfig4}
The partial pair-correlation functions (PCFs) 
for a model structure of {\asih}, with 6,000 
Si atoms and 548 H atoms ($\sim$ 8.4 {\atp} H). 
(Top) A comparison of the simulated Si--Si PCF (blue) 
with the high-energy x-ray diffraction data (red) 
from Laaziri {\etal}~\cite{laaz}
(Bottom) The Si--H PCF for the same model with the 
characteristic first two peaks near 1.51 and 3 {\AA}.
}
\end{figure}

In Table \ref{table1}, we have listed key structural 
properties of {\em ab-initio} relaxed {\asih} models for different sizes 
and hydrogen concentrations. A remarkable feature of 
the models is that they are free from any coordination 
defects and the average value of the bond angles, 
for each size group, is about 109.2{\degree}, which 
is close to the ideal tetrahedral value of 109.47{\degree}. 
The standard deviation of the bond angles ranges from 
8.5 to 10.6{\degree}. We shall see later that these 
values are consistent with those extrapolated from 
Raman measurements. Likewise, the average Si--Si 
bond length for the models is found to be about 2.37 {\AA}, 
which is well within one standard deviation 
of the experimental value of Si--Si  bond 
length of 2.35$\pm$0.065, obtained by Laaziri 
{\etal}~\cite{Laaziri} from high resolution 
x-ray diffraction measurements on annealed/as-implanted 
pure {\asi} samples. A small difference of 0.02 {\AA} can 
be attributed partly due to the use of local 
basis functions in {\sc Siesta} and in part due 
to the presence of H atoms in the network. The 
results for the partial pair-correlation functions, 
$g_{ij}(r)$, for a model with 8.4 {\ath}, are 
presented in Fig.~\ref{nfig4}. The positions 
of the first Si--Si peak at 2.37~{\AA} and the first 
Si--H peak at 1.51~{\AA} in the plots are 
consistent with the results obtained by other 
researchers and experiments~\cite{laaz,street,Dahal}. 

\begin{table}
\caption{A summary of the structural features of the {\asih} models generated in this work. 
Each group is characterized by a base number of Si atoms 
denoted by$N_{\rm Si}$ with a varying H content. For a given group, 
$n_{\rm config}$ is the total number of {\asih} configurations, 
$C_{\rm H}$ is the range of H concentrations in 
the group, 
$r_{\rm Si-Si/H}$ is the range of the Si--Si/H  bond length,
$\langle\theta\rangle$ is the average bond angle, and 
$\Delta \theta$ is the range of the standard deviation of bond angles. 
}
\begin{ruledtabular}
\begin{tabular}{llcccccc}
Group  & $N_{\rm Si}$ & $n_{\rm config}$ & $C_{\rm H}$ & $r_{\rm Si-Si}$ & $r_{\rm Si-H}$ & $\langle\theta\rangle$ & $\Delta \theta$\\
       &              &                  & ({\atp})    &  ({\AA})           & ({\AA})            & (\degree)              & (\degree)   \\
\hline
     1  & 150  & 379  & 6.25-17.58  & 2.36-2.39 & 1.49-1.52 & 109.2 & 8.5-10.4\\
     2  & 200  & 391  & 6.54-18.03  & 2.36-2.39 & 1.50-1.53 & 109.2 & 8.7-10.3\\
     3  & 250  & 378  & 6.02-18.83  & 2.37-2.39 & 1.50-1.52 & 109.2 & 9.0-10.3\\
     4  & 300  & 360  & 6.25-17.58  & 2.37-2.38 & 1.50-1.52 & 109.2 & 8.9-10.4\\
     5  & 350  & 352  & 6.42-18.60  & 2.37-2.38 & 1.50-1.52 & 109.1 & 9.1-10.6\\
     6  & 400  & 329  & 6.10-18.37  & 2.37-2.38 & 1.50-1.52 & 109.2 & 9.1-10.4\\
     7  & 500  & 297  & 6.02-17.22  & 2.37-2.39 & 1.50-1.52 & 109.1 & 9.2-10.4\\
     8  & 1000 & 134  & 6.02-17.76  & 2.37-2.38 & 1.50-1.51 & 109.2 & 9.4-10.4\\
     9  & 3000 & 4    & 7.41-20.5   & 2.41 & 1.52-1.53 & 109.1 & 10\\
     10 & 6000 & 3    & 8.23-8.59   & 2.38 & 1.52-1.53 & 109.1 & 9.9\\
\end{tabular}
\end{ruledtabular}
\label{table1} 
\end{table}

\begin{figure}[tbp!]
\includegraphics[clip,width=0.9\linewidth]{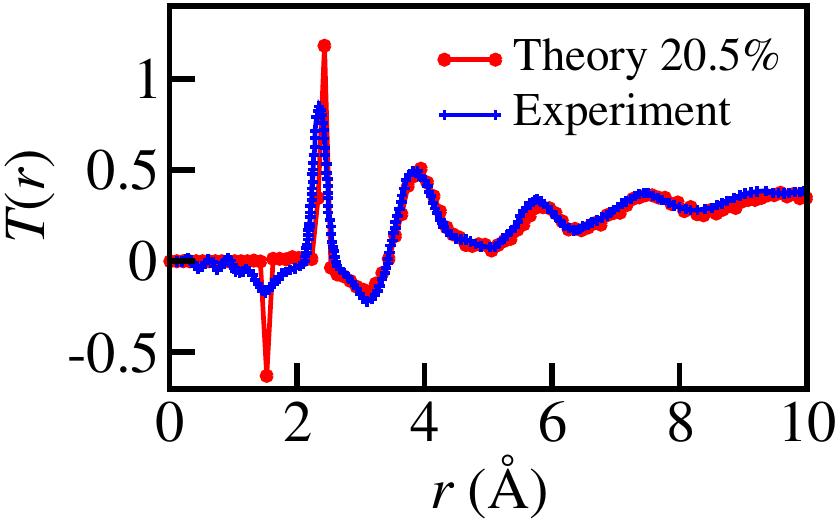}
\caption{\label{fig_corr}
The neutron-weighted total atomic correlation function (red), 
$T(r)$, for a model consisting of 3000 Si atoms with 20.5 {\ath} 
along with its experimental counterpart (blue).
The experimental data for {\asih} correspond to a 
concentration of 22 {\ath} from Ref.~\cite{Wright}. 
}
\end{figure}

In Fig.~\ref{fig_corr}, the total atomic correlation 
function, $T(r)$, of an {\it a}-Si:H structure, consisting 
of 3000 Si atoms, with a concentration of 20.5 
at.\% H is depicted, alongside data from neutron 
scattering experiments for {\asih} samples with 22 {\ath}~\cite{Wright}.  
Here, $T(r)$ is related to the neutron-weighted
conventional atomic pair-correlation function, 
$g(r)$, by $T(r) = 4\pi\rho_0 r g(r)$, and $\rho_0$ 
is the number density of atoms in the network~\cite{tgr}.
The results obtained from our model agree well with those 
from experiments as shown in Fig.~\ref{fig_corr}. 
This establishes that the models accurately produce
the total two-body pair-correlation function as far 
as the radial atomic correlations of {\asih} are 
concerned from experiments.

\begin{figure}[tbp!]
\includegraphics[clip,width=0.9\linewidth]{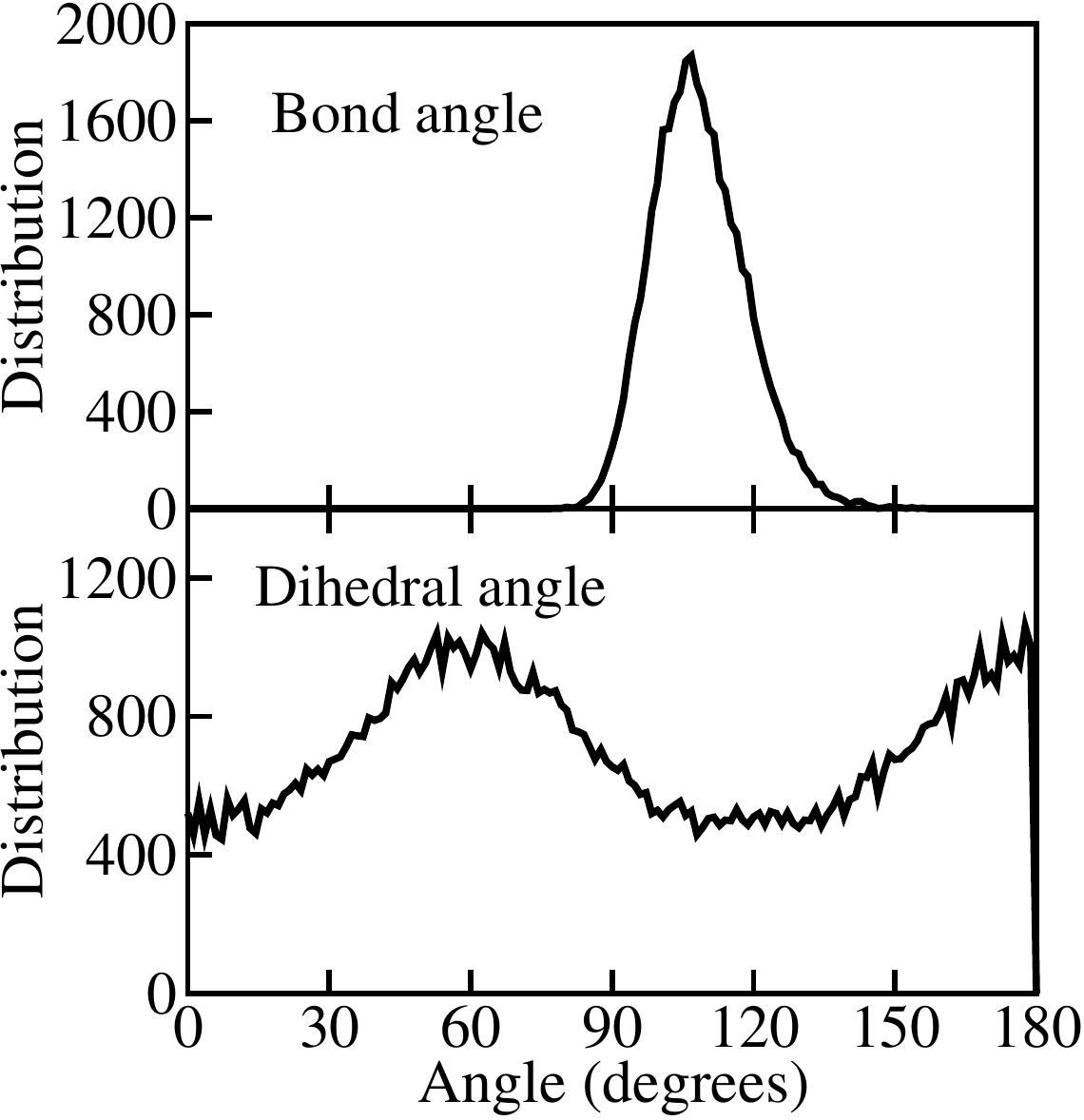}
\caption{\label{fig_angles}
The distribution of bond angles (top) and dihedral 
angles (bottom) for an {\asih} model consisting 
of 6000 Si atoms and 8.2 {\atp} of H. The peak in 
the bond-angle distribution corresponds to a value 
of 109.2{\degree}, with a standard deviation of 
10.4{\degree. The dihedral-angle distribution shows a 
broad peak at 60{\degree}. 
}
}
\end{figure}

Figure~\ref{fig_angles} shows the bond-angle and 
dihedral-angle distributions. The bond-angle 
distribution, which is a measure of the (reduced) 
three-body correlations between atoms, appears 
to be largely Gaussian in its character and it 
has a root-mean-square width of 
$\Delta\theta$=8.5--10.5\degree.
Following Beeman {\it et al.}~\cite{Beeman:1985}, the line 
width of the Raman `optic' peak intensity, $\Gamma$ (in cm$^{-1}$), 
is related to $\Delta\theta$ (in degree) via 
$\Gamma = 15 + 6 \Delta\theta$.  This gives the computed 
values of $\Gamma$ for the models to be in the range 
of 66--79 cm$^{-1}$, which lies well within the range 
of the experimental values of 64--82 cm$^{-1}$, obtained 
from Raman measurements on {\asi}~\cite{Beeman:1985}.
The dihedral-angle distribution, on the other hand, 
provides information concerning the reduced four-body 
atomic correlations, involving a set of four 
consecutive neighboring atoms.  The distribution is 
characterized by the presence of a broad peak in 
the vicinity of 60{\degree} angle. This result is 
unsurprising considering the fact that in 
diamond-structure {\it c}-Si, the angle between two 
dihedral planes is given by 60{\degree}, which 
produces a very sharp delta-function peak in the 
distribution.  The presence of disorder in 
amorphous silicon networks considerably weakens this sharp 
peak, which appears as a bulge at 60{\degree} 
in the dihedral-angle distribution. 

\begin{figure}[tbp!]
\includegraphics[clip,width=0.9\linewidth]{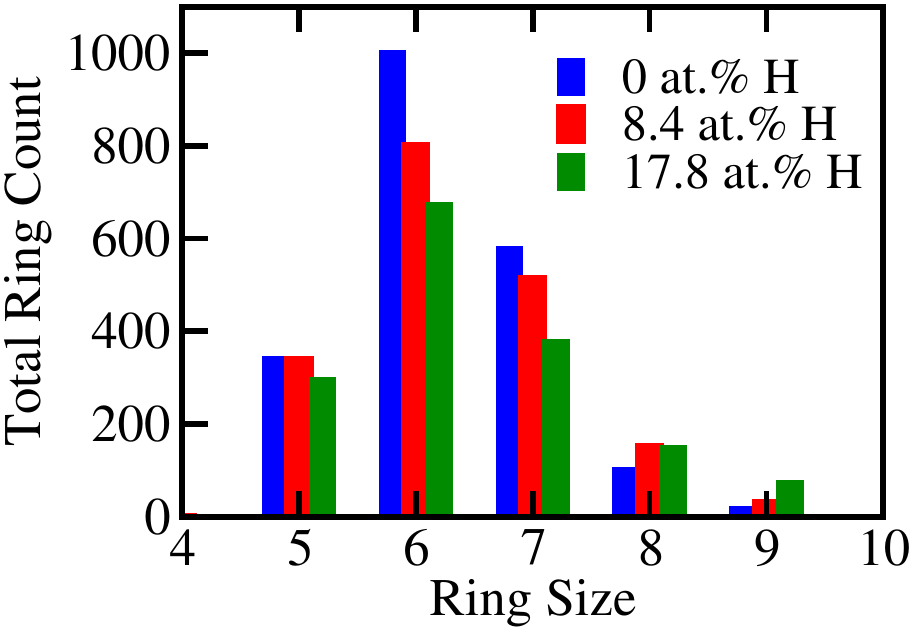}
\caption{\label{fig_rings}
Statistics of total ring counts for three 1000-atom 
Si models with 0, 8.4 and 17.8 {\atp} of H. The 
presence of H atoms leads to a reduction of 5-, 
6- and 7-member rings due to the terminal nature 
of silicon-hydrogen bonding in the amorphous silicon 
environment. 
}
\end{figure}

The structure of tetrahedral amorphous semiconductors 
can be further examined by studying the irreducible 
ring statistics and the distribution of the Voronoi 
volumes of constituent atoms, which provide 
a measure of the network connectivity of the atoms 
and the degree of porosity of the network, respectively. 
By definition, an irreducible ring of size $n$ is the shortest, self-avoiding, 
irreversible closed path or loop, which starts and ends at 
the same atom in $n$ steps. Irreducibility here 
refers to the fact that such a ring cannot be further partitioned 
or reduced into a smaller set of rings without changing the 
topology/connectivity of the path and the dimensionality 
of the amorphous network.
Crystalline silicon in diamond structure is characterized by the 
presence of hexagonal rings only. By contrast, amorphous 
silicon typically has rings of sizes $n$=4--10, with 
6-member rings being statistically dominant. To examine 
how the presence of hydrogen can affect the ring-size 
distribution in {\asi}, we have computed the total number of irreducible 
rings with sizes $n$=3--10 for three 1000-atom models with 
a concentration of 0, 8.4 and 17.8 {\ath}. 
The results are shown in Fig.~\ref{fig_rings} as a histogram. 
As expected, the dominant rings in all the structures are the 
6-member rings. Further, the net ring count is largest for 
{\asi}, which is followed by the structures with 8.4 
and 17.8 {\ath} in the decreasing order. 
The reduction in the ring count for the models with 
8.4 and 17.8 {\ath}, relative to the pure 
amorphous silicon, is due to the presence of too 
many H atoms in the networks that decreases the 
number of available self-avoiding closed loops. 
This is particularly true for Si atoms forming 
a dihydride SiH$_2$ configuration.  A trihydride 
Si atom cannot form a ring in the network. 
It may be noted that none of structures we have studied 
in this work has trigonal or 3-member rings. The presence 
of 3-member rings in {\asi} network is indicative of 
highly strained atomic sites, which often appear in 
poorly produced MD models of {\asi}, quenched from 
the molten state of silicon. High quality models of {\asi} 
and its hydrogenated counterpart should not have any 
3-member rings.

\begin{figure}[ht!]
\includegraphics[clip,width=0.9\linewidth]{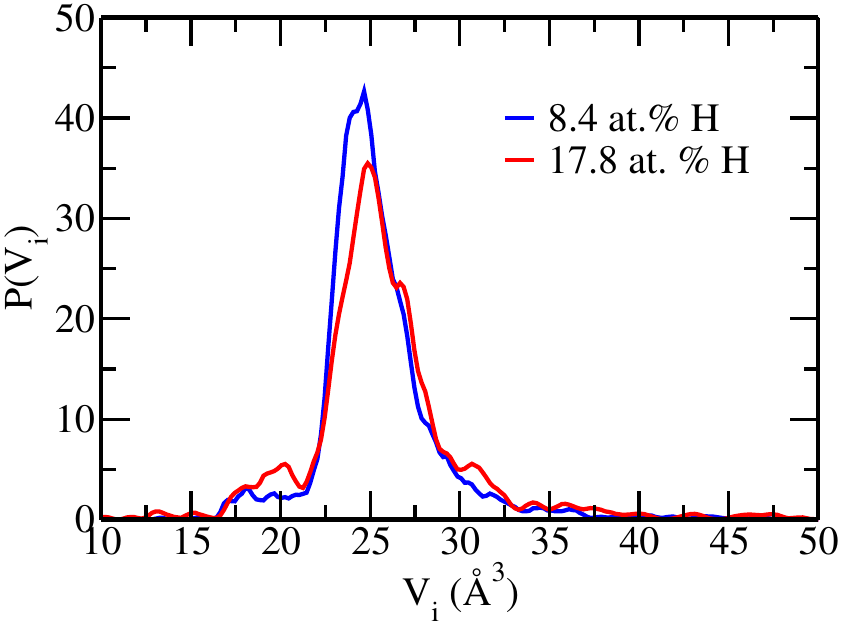}
\caption{\label{voro}
The distributions of the Voronoi volumes ($V_i$) of 
Si atoms obtained from two 1000-atom Si models 
with 8.4 and 17.8 {\atp} of H atoms. The presence 
of more H atoms at the high concentration reduces 
the Voronoi volume of many hydrogen-bonded Si 
atoms, which are reflected in the vicinity of 18--22 
{\AA}$^3$. 
}
\end{figure}

Likewise, the presence of H atoms in {\asih} can 
affect the porosity of the network by reducing its 
effective density with increasing H concentrations. 
This can be readily seen by computing the Voronoi 
volume of the region associated with Si sites.  
The Voronoi volume (of an atom) in a disordered 
network is essentially the analog of the Wigner-Seitz 
cell (of an atom) in a crystal. The Voronoi volume of 
an atom can be calculated from the positions of the atom 
and its nearest neighbors~\cite{vnote}. 
The formation of various silicon-hydrogen bonding 
configurations at high concentration of H can lead 
to a greater number of silicon monohydrides and 
dihydrides, which in turn reduce the Voronoi volume 
of Si atoms that are bonded to H atoms. This is 
apparent from Fig.~\ref{voro}, 
where the distributions of Voronoi volumes, $P(V_i)$ 
versus $V_i$, for two models with 8.4 and 17.8 {\ath}, 
are plotted for comparison. The results correspond 
to the configuration averaged values of $V_i$, 
at silicon site $i$, obtained from two independent 
configurations.  The primary peak at 25 {\AA}$^3$ 
corresponds to those Si sites that are bonded to 
four Si neighbors, whereas the secondary peak 
near 20 {\AA}$^3$ arises from the hydrogen-bonded 
Si atoms in the network. The height of the 
secondary peak thus increases with the increasing 
number of H atoms at high concentrations. 

\subsection{Hydrogen microstructure and NMR spectra of {\asih}} 

The distribution of hydrogen in {\asi} plays an 
important role in determining electronic and 
optical properties of {\asih}. Since pure
samples of {\asi} almost always contain some 
coordination defects, mostly in the form of 
vacancy-type defects and a few dangling 
bonds, it is necessary to eliminate 
these defects via hydrogen passivation to obtain 
device-grade samples of {\asih} for technological 
applications. Hydrogen can also break weak 
Si--Si bonds to form energetically more stable 
SiH/SiH$_2$ bonding configurations, which minimize 
local network strain. The presence of weak Si--Si bonds 
is related to the degree of disorder in 
the network, which depends on experimental 
methods and preparation conditions used to 
produce the samples. A similar observation 
applies to {\asih} models obtained from 
different simulation techniques. 

In general, hydrogenation of {\asi} leads to a 
formation of silicon monohydride (SiH) and 
dihydride (SiH$_2$) configurations at low concentrations 
of 6--10~{\atp} H. The resulting samples/models 
are often described as device-grade quality. Further addition 
of H atoms in the network can result in a highly 
clustered environment of SiH/SiH$_2$ bonds, a 
few SiH$_3$ configurations, and molecular 
hydrogen (H$_2$) formation inside small cavities 
at high concentrations of 15--20 {\atp} H~\cite{Biswas2017,Chabal1984}. 
These aspects of concentration dependent hydrogen 
microstructure can be studied experimentally by 
using an array of techniques, such as infrared (IR) 
spectroscopy~\cite{Chabal1984}, nuclear 
magnetic resonance (NMR)~\cite{nmr1, nmr2} 
and positron-annihilation spectroscopy (PAS)~\cite{Sekimoto2016}.  
Likewise, the hydrogen microstructure of {\asih} models 
can be computationally studied by analyzing hydrogen 
distributions in the network with emphasis on the 
presence of SiH, SiH$_2$ and SiH$_3$ configurations, 
and the spatial distribution of hydrogen at low 
and high concentrations. The line shape of the 
NMR spectra of {\asih} provides useful information 
on the distribution of H atoms in the network. 
\begin{table}[t!] 
\caption{\label{table2}{
Hydrogen microstructure at high and low H concentrations. 
$N_{\rm Si}$ and $C_{\rm H}$ indicate the number of 
Si atoms and the range of H concentration in at.~\%. 
SiH$_n$ denotes the number of H atoms (in at.~\%) present 
in the networks 
as silicon monohydrides, dihydrides, and trihydrides, 
for $n$=1, 2, and 3, respectively.  The numbers of 
isolated, sparse, and clustered H atoms are also 
listed here. The results were obtained by averaging 
over 10 independent configurations for each 
concentration. 
}
}
\begin{ruledtabular}
\begin{tabular}{lccccccc}
$N_{\rm Si}$ & $C_{\rm H}$ & SiH    & SiH$_2$ & SiH$_3$ & Isolated & Sparse & Clustered\\
             &  (\%)       & (\%)   & (\%)    & (\%)    &  (\%)    & (\%)   & (\%) \\
\hline
500          & 6.0-7.8         & 80.6   & 19.4    & 0       & 14.6     &  55.8    & 29.6 \\
500          & 14.1-17.2       & 79.9   & 18.6    & 1.5     & 1.4      &  45.7    & 52.9 \\
1000         & 6.2-8.8         & 82.6   & 16.2    & 1.2       & 15.3     &  52.7    & 32.3 \\
1000         & 15.5-17.8       & 82.2   & 15.9    & 1.9     & 2.2      &  43.9    & 53.9 \\
\end{tabular}
\end{ruledtabular}
\end{table} 

\begin{figure}[ht!]
\includegraphics[clip,width=0.6\linewidth]{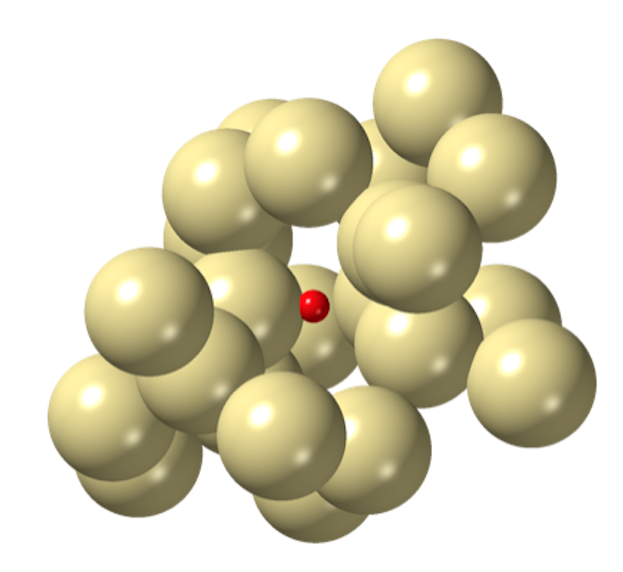}
\caption{
\label{isoH}
An isolated H atom (red) in an {\asih} configuration with 1000 
Si atoms (yellow), with a hydrogen concentration of 
{17.8 {\atp}} H.  There are no H atoms within a sphere of 
radius 5 {\AA}, except the lone central H atom.  
}
\end{figure} 

\begin{figure}[ht!]
\includegraphics[clip,width=0.6\linewidth]{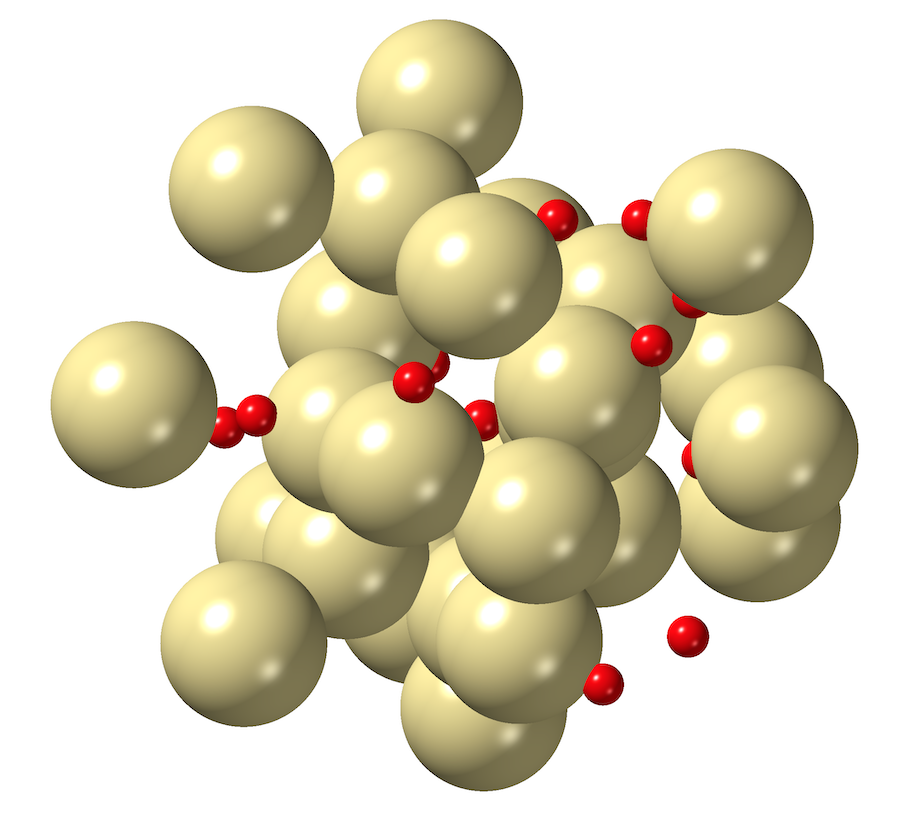}
\caption{
\label{clusH}
A clustered distribution of H atoms (red) observed in 
an {\asih} configuration with 1000 Si atoms (yellow), 
with a hydrogen concentration of {17.8 {\atp}}.  
The spherical region shown above corresponds to a 
radius of 5 {\AA} around a chosen central H atom with 
additional 11 H atoms within the region.  
}
\end{figure} 

Table \ref{table2} lists some characteristic properties 
of microstructural distributions of H atoms at low and 
high concentrations of 6--8.8 {\atp} H and 
14.1--17.8 {\atp} H, respectively. 
The values listed in the table were obtained by averaging 
over 10 structures per group.  
It is apparent from the table that the great majority 
(about 80 {\atp}) of H atoms 
reside in the networks as silicon monohydrides (SiH), 
which are followed by 16--20 {\atp} of H atoms as silicon 
dihydrides (SiH$_2$). A few SiH$_3$ configurations 
(no more than 2\%) are also found to be present in the 
networks at high concentrations.  However, none of 
the models in our study shows the presence of any 
H$_2$ molecules even at the high concentration of 
18 {\ath}.  We surmise that this is due to the absence of voids 
in model {\asih} networks. It has been shown 
elsewhere~\cite{Chabal1984,Biswas2017,Biswas2021,Sekimoto2016} 
that the presence of voids plays an important role in 
the formation of non-bonded hydrogen, such as H$_2$ 
molecules, in {\asih}.
The number of SiH$_n$ configurations (with $n$=1, 2, 3) 
and their spatial distributions in the networks 
constitute a description of hydrogen microstructure 
in {\asih}. 

\begin{figure}[t!]
\includegraphics[clip,width=0.65\linewidth]{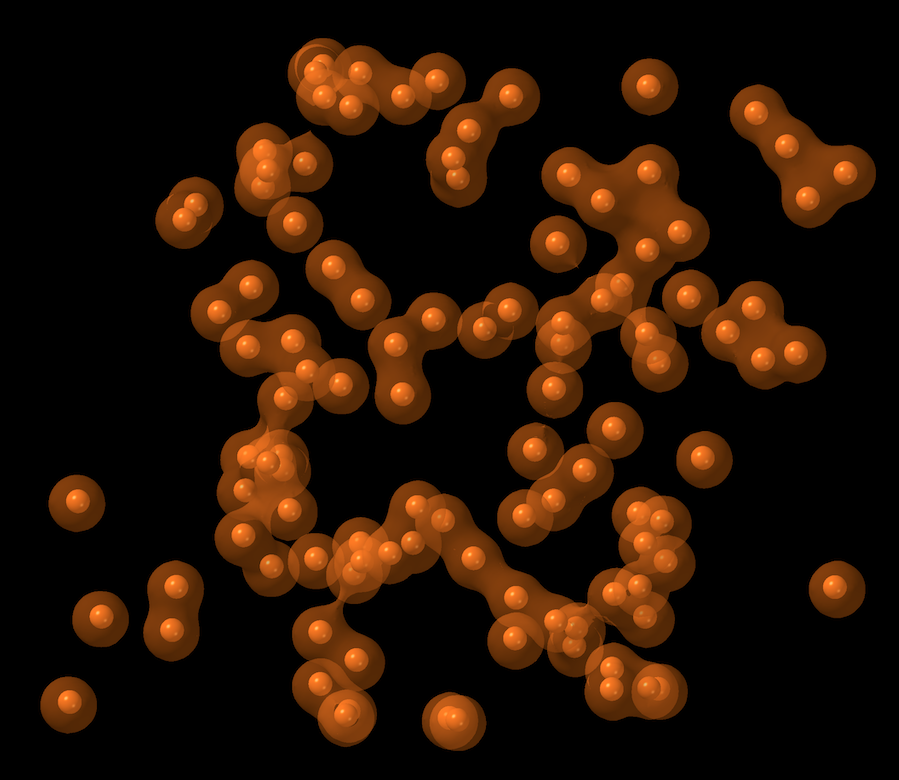}
\caption{
\label{sparse1000}
Hydrogen microstructure in {\asih} showing a relatively sparse 
distribution of H atoms (orange) in a configuration with 
1000 Si atoms and 17.8 {\ath}. Clustered H atoms are 
not shown in the figure for clarity.  The radius of 
the translucent outer spheres surrounding H atoms 
corresponds to a value of 1.5 {\AA}, indicating 
distances between H atoms in the vicinity. Silicon atoms 
are not shown in this figure.
}
\end{figure}

A further characterization of hydrogen microstructure 
in {\asih} can be made from the distribution of 
H atoms within the networks.  The spatial distribution of 
hydrogen in {\asih} can be characterized by examining 
the vicinity and abundance of H atoms, and classifying 
the distribution as a sparse or clustered distribution.
Following Baum {\etal}~\cite{Baum}, a hydrogen 
atom is considered to be 
isolated if there are no neighboring hydrogen atoms within a sphere 
of radius of 4--5 {\AA}. 
Likewise, a hydrogen atom is assumed to be a part 
of a cluster if the atom has at least 6--8 H 
neighbors within a radius of 4--5 {\AA}. The 
remaining H atoms may be counted as sparsely 
distributed in the network~\cite{sparse}. 
Figures~\ref{isoH} and \ref{clusH} show a representative example 
of an isolated H atom and a clustered 
distribution of H atoms, respectively, 
within a spherical region of radius  5 {\AA}, 
in an {\asih} network with 17.8 {\atp} of H. 
Similarly, the microstructure originated 
from a sparse distribution of H atoms in 
an {\asih} network with 17.8 {\atp} H 
is presented in Fig.~\ref{sparse1000}.  
As shown in Fig.~\ref{sparse1000}, the minimum 
center-to-center distance between any two H 
atoms is found to be at least 3 {\AA}.  These 
sparsely distributed H atoms neither belong 
to any H clusters nor do they satisfy the 
criterion for being isolated H atoms.  
Table \ref{table2} lists the number of H atoms 
distributed in isolated, sparse and clustered 
environments of {\asih} in a few representative 
models.  Here, the results correspond to a (cluster) 
radius of 4 {\AA} and the minimum size of H 
clusters is chosen to be 6, which includes the 
central H atom. The choice of these values was 
guided by the experimental results from NMR 
measurements~\cite{nmr1, nmr2,Baum}. 
We shall see soon that these 
sparsely distributed H atoms and H clusters 
play an important role in determining the 
shape of the NMR line spectra of {\asih}. 
Together with Table \ref{table2}, Figs.~\ref{isoH} 
to \ref{sparse1000} provide a full description 
of hydrogen microstructure at low and high 
concentrations of H atoms in {\asih} networks. 
\begin{figure}[t!]
\includegraphics[clip,width=0.8\linewidth]{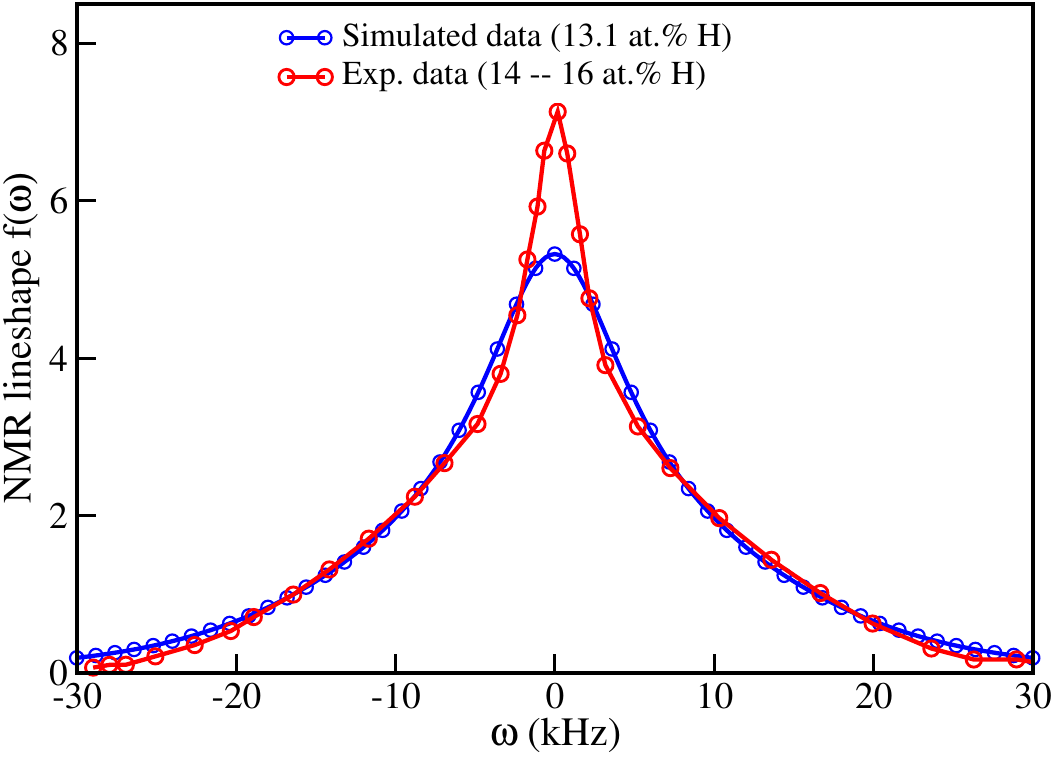}
\caption{
\label{nmr3000}
A comparison of the simulated (blue) NMR line shape 
function, $f(\omega)$, obtained from moment-based 
considerations, with the experimental data 
(red) from Ref.~\onlinecite{Taylor}. The simulated 
data correspond to a 3000-atom model with 13.1 {\ath}, 
whereas the experimental data are from {\asih} samples 
with about 14 {\ath}. 
}
\end{figure}

Experimental data from NMR measurements~\cite{Taylor,nmr1,nmr2} 
indicate that microstructural distributions of H atoms 
in {\asih} networks can play a considerable role 
in forming the shape of the resonance curve~\cite{Taylor,nmr1,nmr2}. 
In particular, the shape of the resonance curve can provide 
an indirect means to further characterize the distribution of 
H atoms in {\asih} networks.  However, a direct 
calculation of the resonance curve is rather 
complicated, involving quantum-mechanical 
manybody calculations, and outside the scope of 
the present study. We thus resort to a simpler 
approach to construct an approximate form of the resonance 
curve from moment-based considerations, as discussed 
by Van Vleck~\cite{Vleck} and others~\cite{Abragambook,biswas-nmr}. 
In this approach, one assumes that the resonance curve 
can be approximated as a linear combination of a 
suitable Gaussian function and a truncated Lorentzian, 
which are weighted by the concentration of 
clustered H atoms and sparsely distributed H atoms 
(including isolated ones) present in a network, 
respectively. The approximation 
relies on theoretical and experimental observations 
that the dipolar interaction between H atoms at 
low concentrations produces a narrow Lorentzian-like 
curve, whereas the presence of small H clusters at 
high concentrations leads to a broad Gaussian shape 
of the resonance curve. 
A detailed theoretical analysis of this 
approach is given by Van Vleck~\cite{Vleck}. 
Following this approach, we have constructed the 
resonance curve for a 3000-atom Si model with 
13.1 {\ath}. The choice of a large model is 
necessary for this purpose in order to 
produce a statistically robust value of the 
first two moments of the hydrogen distribution 
in the network.  Figure~\ref{nmr3000} shows 
the approximate shape of the resonance curve, 
$f(\omega)$, along with the same from experiments. 
A comparison of our results with the experimental data 
from a sample with 14 {\atp} H in Ref.~\onlinecite{Taylor} 
indicates that the moment-based approach 
can indeed reproduce the approximate line shape 
of the experimental resonance curve of {\asih}.

\subsection{Electronic and optical properties of {\asih}}

Although structural properties and the microstructural distribution 
of H atoms in {\asih} networks play a crucial role in 
determining the quality of the models, the most 
definitive test follows from the electronic properties 
of the models. The electronic quality of a model is 
chiefly determined by the electronic density 
of states (EDOS) and the size of the electronic band gap. The 
latter depends on a number of factors, such as the 
degree of bond-length and bond-angle disorder that 
affect the states near the edges of the band gap, 
also known as the Urbach tails, and in 
particular the presence of dangling (and possibly 
floating) bonds in the networks, leading to midgap and tail states. 
Furthermore, the application of the local density 
approximation (LDA) of the exchange-correlation (XC) 
energy in density functional theory is also known 
to underestimate the size of the gap.

All {\asih} models reported in this study exhibit 
a clean gap in the corresponding EDOS. This 
is not surprising as the models are characterized 
by a very narrow bond-angle distribution (see 
Table \ref{table1}) and silicon atoms therein 
are all 100\% four-fold-coordinated, 
with no defects in the networks. The size of the electronic gap can be 
estimated by computing the energy eigenvalue 
difference between the highest occupied molecular orbital (HOMO) and the 
lowest unoccupied molecular orbital (LUMO), also 
known as the HOMO-LUMO gap, which lies in the 
range of 1--1.8 eV for the models studied in this 
work. 

\begin{figure}[tbp!]
\includegraphics[clip,width=0.8\linewidth]{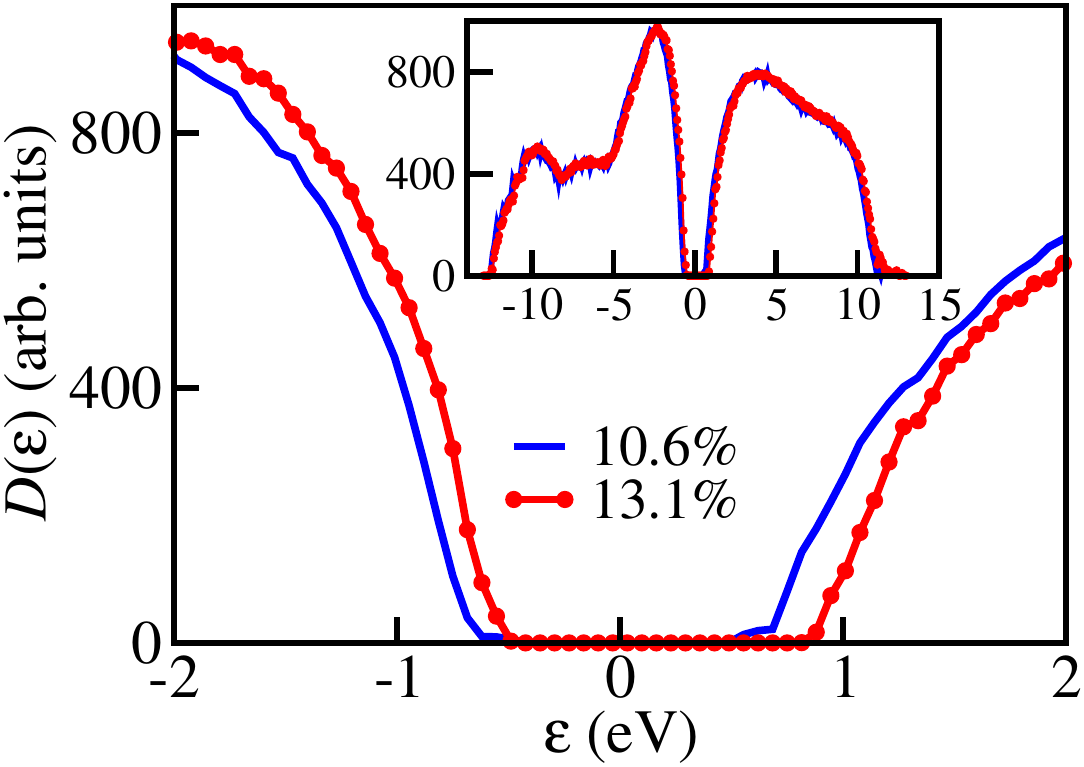}
\caption{
\label{edos}
EDOS plots for two {\asih} configurations with 
hydrogen concentrations of 10.6 (blue) and 13.1 
(red) {\ath}. 
The models have produced a pristine electronic gap, and 
the size of the gap can be seen to have slightly 
increased with the concentration of H atoms in 
the models, as observed in experiments.  The 
full electronic density of states is shown in 
the inset.
}
\end{figure}

The band-gap regions of the EDOS of two representative 
{\asih} models with 3000 Si atoms are plotted in 
Fig.~\ref{edos}, along with the full EDOS as an 
inset. 
The concentrations of H atoms in the models are given by 
10.7 and 13.1 {\atp}, which correspond to 358 and 452 H 
atoms in the networks, respectively. The EDOS is obtained 
by writing $D(\varepsilon)=\sum_i \delta(\varepsilon-\varepsilon_i)$, 
where $\{\varepsilon_i\}$ are the single-particle 
Kohn-Sham eigenvalues, and convoluting each $\delta$
function with a Gaussian distribution centered at 
$\varepsilon_i$. The width of the Gaussian distribution 
is so chosen that it is a fraction of the average 
level-spacing distance between two neighboring eigenvalues. 
It is evident from Fig.~\ref{edos} that the models produced 
a pristine electronic gap with no defect-induced 
localized midgap states. The size of the gap can be seen to be 
somewhat dependent on the hydrogen content of the 
models; the presence of more hydrogen leads to 
a slightly larger gap in Fig.~\ref{edos}. The widening 
of the band gap with increasing H concentrations 
can be attributed to the replacement of weak Si--Si 
bonds by energetically stable Si--H bonds, which 
reduce the overall strain and local disordering in a 
network and some of the (tail) states in the 
vicinity of the gap.  

\begin{figure}[tbp!]
\includegraphics[clip,width=0.8\linewidth]{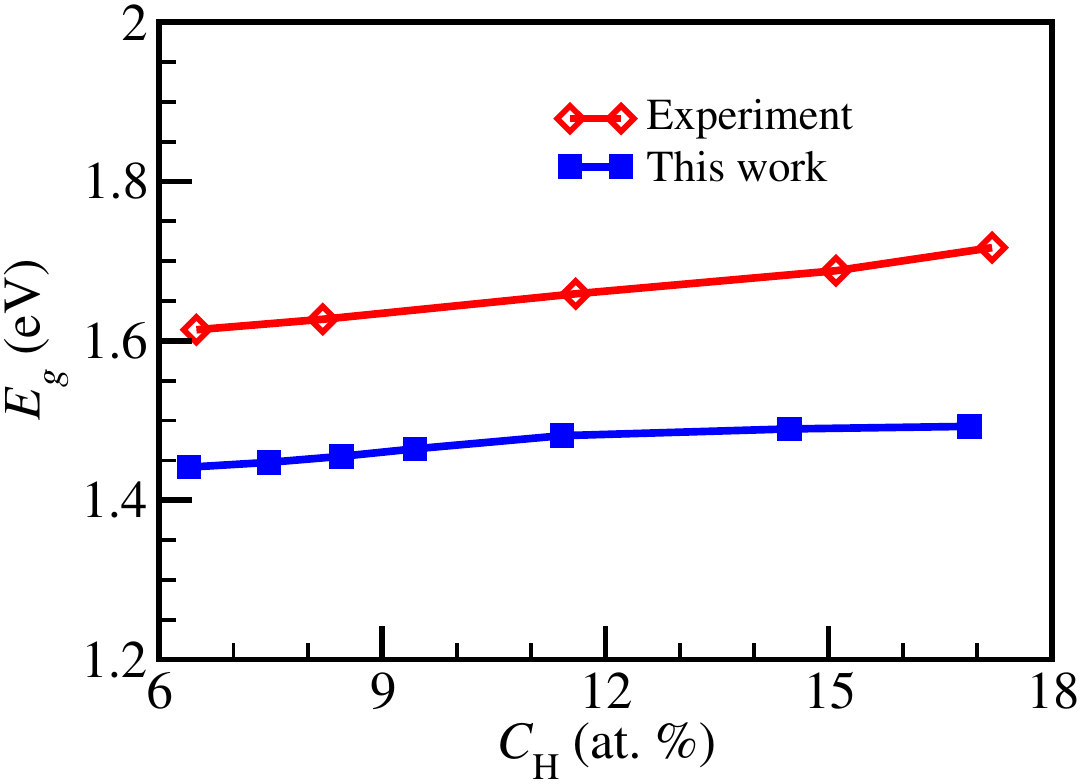}
\caption{\label{fig_gap}
The dependence of the Tauc optical gap ($E_g$) with the 
concentration of hydrogen ($C_H$) from 6 to 17 {\atp}. 
The corresponding experimental data (red diamond) 
shown in the plot are from Kageyama {\it et al.}~\cite{Kageyama}
}
\end{figure}

In order to study the variation of the band gap with 
hydrogen concentrations and to compare the results 
with experimental data, it is more useful to calculate 
the optical gap of a model. This can be achieved 
by using the Tauc model~\cite{Tauc:1966,Tauc:1968}, which 
is based on the following 
assumptions: (i) The electronic density of states in 
the vicinity of the valence band edge and the 
conduction band edge exhibit a parabolic 
behavior; (ii) The wavevector ($\vec{k}$) selection 
rules can be relaxed as $\vec{k}$ is no longer a 
good quantum number in noncrystalline solids; 
(iii) The weak energy dependence of the transition 
matrix elements between the valence and conduction 
band edge states can be ignored.  To calculate the 
optical gap, one writes the integrated joint density 
of states~\cite{morigaki} 
\be
\mathcal{J}(\hbar\omega)=\int D_{_{\rm V}}(\varepsilon)
D_{_{\rm C}}(\varepsilon + \hbar\omega) \, d\varepsilon, 
\label{jdos}
\ee
where $D_{_{\rm V}}$ and $D_{_{\rm C}}$ are the 
valence and conduction band electronic densities 
of states, respectively, and the integration is 
carried out over all pairs of states in the valence 
and conduction bands that are energetically separated 
by $\hbar\omega$.  The foregoing assumptions lead to 
the following simplified expression of Eq.~(\ref{jdos})
\be
\sqrt{\mathcal{J}(\hbar\omega)} = 
\sqrt{\alpha \hbar \omega} = M_{_{\rm Tauc}}(\hbar\omega - E_{g}). 
\label{Tauc1}
\ee
\noindent 
Here, $\alpha$ is the absorption coefficient, 
$M_{_{\rm Tauc}}$ is a positive constant, 
and $E_{g}$ is the so-called Tauc optical 
gap. Equation (\ref{Tauc1}) suggests that 
the value of $E_{g}$ can be extrapolated from the 
plot of $\sqrt{\mathcal{J}(\hbar\omega)}$ 
versus $\hbar\omega$.  
Figure~\ref{fig_gap} shows the plot of the average 
optical gap $E_g$ with increasing H concentration $C_H$. 
The averaging was done as follows: 
(i) all 2,627 optical gaps were divided into seven 
concentration groups, with each group spanning a hydrogen
concentration range; (ii) the gaps within each group 
were averaged to yield a single gap $E_g$ and the 
corresponding H concentrations within the group 
were averaged to yield a single concentration $C_H$.
The experimental data obtained by Kageyama 
{\etal}~\cite{Kageyama} are also plotted in Fig.~\ref{fig_gap} 
for comparison.  It is apparent from the plot that 
the calculated values of $E_g$ show an approximate 
linear behavior with the concentration of H atoms 
from 6 to 17 {\atp}. This behavior is found to be 
very consistent with the experimental results obtained 
by Kageyama {\etal}~\cite{Kageyama} It goes without 
saying that the calculated values of the gap are 
somewhat smaller than the experimental values. This 
is partly due to the use of the LDA of the 
exchange-correlation functional in DFT calculations 
and in part due to the somewhat limited size of the 
basis functions employed in our calculations. 

\subsection{Vibrational properties of {\asih}}

To further characterize the {\asih} models, one often 
computes the vibrational density of states (VDOS) 
and compares the results with those from inelastic 
neutron-scattering experiments. In solids, the 
atomic vibrational energy scale is much 
smaller than its electronic counterpart, and 
therefore VDOS can provide a finer aspect of 
the local atomic environment of Si and H atoms, 
involving atomic vibrations in a disordered 
environment, which often may not be apparent in 
structural and electronic properties. 

\begin{figure}[tbp!]
\includegraphics[clip,width=0.8\linewidth]{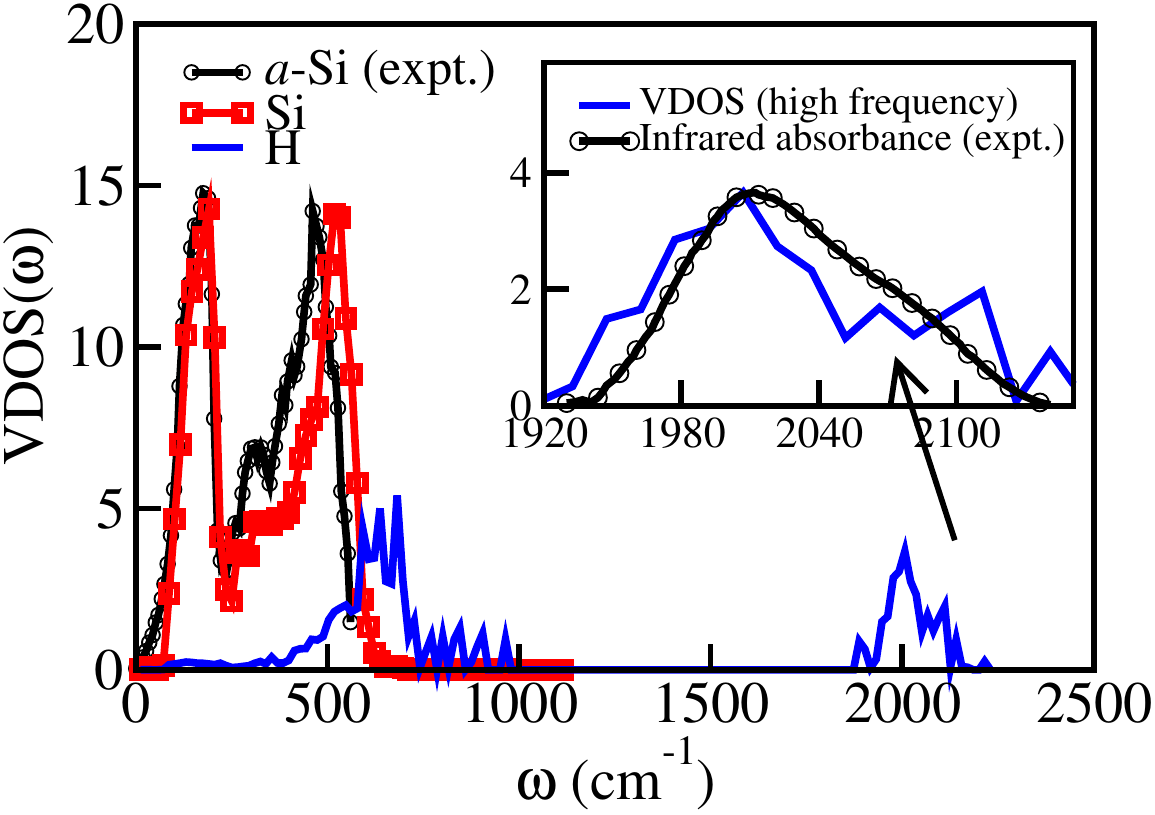}
\caption{\label{sih_vdos_fig} 
The vibrational density of states (VDOS) for 
a 556-atom model of {\asih} with 10.4 {\ath}. 
The VDOS projected at Si sites (red) 
and H sites (blue) are shown, along with the 
experimental data (black) for {\asi} from neutron scattering 
measurements~\cite{Kamita}. 
The inset shows a comparison of the high-frequency 
region of the VDOS with the experimental infrared 
absorbance data obtained by Scharff {\it et al}.~\cite{Scharff}
} 
\end{figure} 

The VDOS of {\asih} was computed using the harmonic 
approximation. In the harmonic approximation, the 
contribution to the total energy of a solid, 
upon a small displacement from the equilibrium 
configuration, from cubic and higher-order terms 
are neglected. Following the theory of small 
oscillations, one can readily construct a 
force-constant matrix by displacing one atom at 
a time along six coordinate directions ($\pm x$, 
$\pm y$, $\pm z$). To preserve the harmonic 
character of atomic vibrations, the displacement 
of the atoms is typically restricted to 
0.005--0.01 {\AA}. The diagonalization of the mass-adjusted 
force-constant matrix, also known as the dynamical 
matrix (DM), then yields the squared eigenfrequencies, 
$\omega_i^2\;(i=1,2,\cdots\,3N)$, and the normalized 
eigenvectors, $\{{\bf \Phi}^j(\omega_i):\,j=1,2,\cdots,N;\, 
i=1,2,\cdots\,3N\}$, where ${\bf \Phi}^j(\omega_i)
=[\Phi^j_x(\omega_i), \Phi^j_y(\omega_i), 
\Phi^j_z(\omega_i)]$. 
The VDOS was computed as $g(\omega) =\sum_{i=1}^{3N}\delta(\omega-\omega_i)$ 
and the inverse participation ratio (IPR) of each (normalized) 
vibrational mode as ${\rm IPR} (\omega_i)=\sum_{j=1}^{3N}|{\bf \Phi^j_i}|^4$. 
The IPR of a normal mode provides a simple measure of the 
degree of amplitude localization of the mode in 
real space. A value of IPR$\sim$1 suggests a highly 
localized mode around a site, whereas IPR$\sim 1/N$ 
indicates a delocalized mode over $N$ sites.  
The VDOS and IPR calculations were carried out 
using a 556-atom {\asih} model with 10.4 {\atp} 
of H, and ten independent configurations were 
used to obtain the configurationally averaged 
values of the VDOS. 

\begin{figure}[tbp!]
\includegraphics[clip,width=0.8\linewidth]{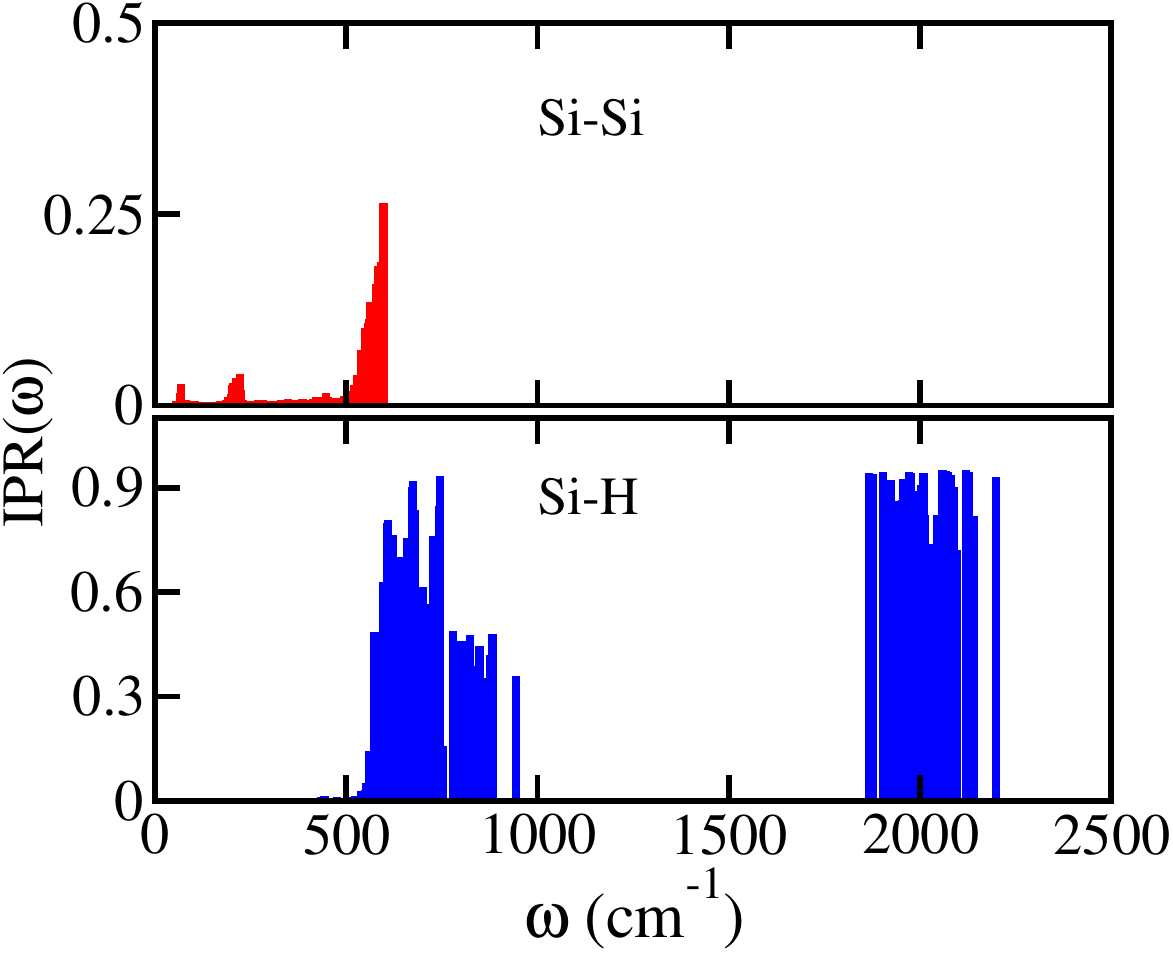}
\caption{\label{IPR_fig} 
Inverse participation ratios (IPR) of the vibrational 
normal modes of a 556-atom model with 10.4 {\ath} 
for different mode frequencies. The upper 
panel corresponds to the results where the modes 
are associated with the vibrations of Si atoms 
in pure Si environment.  The lower panel shows 
the results for Si atoms bonded to H atoms below 
1000 cm$^{-1}$ and high-frequency localized 
modes for H atoms in the vicinity of 2000 cm$^{-1}$. 
A high IPR value corresponds to a localized mode.
} 
\end{figure} 

In Figs.~\ref{sih_vdos_fig} and \ref{IPR_fig}, we 
depict the plots of the VDOS (decomposed into Si 
and H contributions) and IPR (decomposed in Si--Si 
and Si--H vibrations), respectively. The VDOS 
for Si atoms is compared with the results obtained 
from inelastic neutron scattering measurements on 
{\asi} and the infrared absorbance data for {\asih} 
from Refs.~\onlinecite{Kamita} and 
\onlinecite{Scharff}, respectively.
The VDOS in Fig.~\ref{sih_vdos_fig} shows three 
well-defined bands: (i) $\omega < 600 \; 
{\rm cm}^{-1}$; (ii) $500 < \omega < 900 \; 
{\rm cm}^{-1}$; (iii) $1900 < \omega < 
2200\;{\rm cm}^{-1}$. 
The low-frequency band mainly comprises delocalized Si--Si modes. 
The delocalized nature of the modes is evidenced by small 
values of the IPR (red bars) in Fig.~\ref{IPR_fig}. 
The mid-frequency band essentially 
consists of Si--H wagging modes in the frequency range of 
630--660 ${\rm cm}^{-1}$ and Si--H bending modes in the 
830--860 ${\rm cm}^{-1}$ range. Experimental observations 
in Ref.~\onlinecite{Kamita} suggest that the wagging and 
bending modes primarily originate from SiH$_2$ configurations. 
An analysis of IPRs and eigenvectors of the DM indeed 
confirmed that the vibrational amplitudes from SiH$_2$ 
units dominate the mid-frequency band. By contrast, 
the high-frequency band is found to be composed of 
Si--H stretching modes within the frequency range 
of 2000--2200 ${\rm cm}^{-1}$.  The Si--H modes in 
the mid- and high-frequency bands are found to be 
highly localized on H atoms from eigenvector analyses. 
This observation is also corroborated by IPR values 
for the respective modes, which are very close to unity 
in most of the cases [See Fig.~\ref{IPR_fig}\,(lower panel)].

\section{Web site for structural database}\label{website}
A major outcome of the present study is the development 
of a large structural database of {\asih}. The study 
has resulted more than two thousand defect-free, high-quality 
atomic configurations of {\asih}. The database consists 
of atomic configurations of size from 150 to 6000 Si 
atoms with H concentrations in the range from 6 to 
19 {\atp} H. A provisional database, with 
several hundreds of {\asih} configurations, is currently 
up and functional, which lists the basic structural and 
electronic properties of the {\asih} models.
The atomic coordinates and associated properties of 
the {\asih} configurations are downloadable upon 
registration from the provisional database located at 
\href{https://drive.google.com/drive/folders/1JFjI5W0wrgkhxjzShINHk1tuwIr3zmh7?usp=sharing}{www.drive.google.com/SiH-models}. 
The complete database will be soon available and hosted 
at a permanent location in the near future.  Atomic 
configurations and related metadata will be freely 
available to the public for academic research on 
request. 

\section{Conclusions}\label{conclusion}

A computationally efficient method for atomistic 
modeling of hydrogenated amorphous silicon has been presented. 
A remarkable feature of the method is that it can produce large 
realistic structures of {\asih} with a variety of 
silicon-hydrogen bonding configurations, which are 
consistent with the associated hydrogen 
microstructure for varying hydrogen concentrations 
observed in experiments. 
The method combines the power of classical 
metadynamics simulations with density functional 
theory in generating a large class of models 
with varying hydrogen contents.  
Using the method, we have built a large structural 
database of {\asih}, consisting of 2,600+ independent 
{\asih} configurations, with atomic hydrogen 
concentrations in the range from 6 to 20 {\ath}. 
All structures are high quality 
in the sense that they are (a) free from any 
coordination defects and (b) the structural, electronic, 
vibrational, and microstructural properties of 
hydrogen distribution in the structures are in 
good agreement with experiments. The structural 
database resulted from this work will provide 
a direct support to the materials modeling community 
in studying problems involving the physics of 
{\asih} and {\asih}-based devices. 


\begin{thebibliography}{69}%
\makeatletter
\providecommand \@ifxundefined [1]{%
 \@ifx{#1\undefined}
}%
\providecommand \@ifnum [1]{%
 \ifnum #1\expandafter \@firstoftwo
 \else \expandafter \@secondoftwo
 \fi
}%
\providecommand \@ifx [1]{%
 \ifx #1\expandafter \@firstoftwo
 \else \expandafter \@secondoftwo
 \fi
}%
\providecommand \natexlab [1]{#1}%
\providecommand \enquote  [1]{``#1''}%
\providecommand \bibnamefont  [1]{#1}%
\providecommand \bibfnamefont [1]{#1}%
\providecommand \citenamefont [1]{#1}%
\providecommand \href@noop [0]{\@secondoftwo}%
\providecommand \href [0]{\begingroup \@sanitize@url \@href}%
\providecommand \@href[1]{\@@startlink{#1}\@@href}%
\providecommand \@@href[1]{\endgroup#1\@@endlink}%
\providecommand \@sanitize@url [0]{\catcode `\\12\catcode `\$12\catcode
  `\&12\catcode `\#12\catcode `\^12\catcode `\_12\catcode `\%12\relax}%
\providecommand \@@startlink[1]{}%
\providecommand \@@endlink[0]{}%
\providecommand \url  [0]{\begingroup\@sanitize@url \@url }%
\providecommand \@url [1]{\endgroup\@href {#1}{\urlprefix }}%
\providecommand \urlprefix  [0]{URL }%
\providecommand \Eprint [0]{\href }%
\providecommand \doibase [0]{https://doi.org/}%
\providecommand \selectlanguage [0]{\@gobble}%
\providecommand \bibinfo  [0]{\@secondoftwo}%
\providecommand \bibfield  [0]{\@secondoftwo}%
\providecommand \translation [1]{[#1]}%
\providecommand \BibitemOpen [0]{}%
\providecommand \bibitemStop [0]{}%
\providecommand \bibitemNoStop [0]{.\EOS\space}%
\providecommand \EOS [0]{\spacefactor3000\relax}%
\providecommand \BibitemShut  [1]{\csname bibitem#1\endcsname}%
\let\auto@bib@innerbib\@empty
\bibitem [{\citenamefont {Taguchi}\ \emph {et~al.}(2014)\citenamefont
  {Taguchi}, \citenamefont {Yano}, \citenamefont {Tohoda}, \citenamefont
  {Matsuyama}, \citenamefont {Nakamura}, \citenamefont {Nishiwaki},
  \citenamefont {Fujita},\ and\ \citenamefont {Maruyama}}]{HIT1}%
  \BibitemOpen
  \bibfield  {author} {\bibinfo {author} {\bibfnamefont {M.}~\bibnamefont
  {Taguchi}}, \bibinfo {author} {\bibfnamefont {A.}~\bibnamefont {Yano}},
  \bibinfo {author} {\bibfnamefont {S.}~\bibnamefont {Tohoda}}, \bibinfo
  {author} {\bibfnamefont {K.}~\bibnamefont {Matsuyama}}, \bibinfo {author}
  {\bibfnamefont {Y.}~\bibnamefont {Nakamura}}, \bibinfo {author}
  {\bibfnamefont {T.}~\bibnamefont {Nishiwaki}}, \bibinfo {author}
  {\bibfnamefont {K.}~\bibnamefont {Fujita}},\ and\ \bibinfo {author}
  {\bibfnamefont {E.}~\bibnamefont {Maruyama}},\ }\bibfield  {title} {\bibinfo
  {title} {Record efficiency hit solar cell on thin silicon wafer},\ }\href
  {https://doi.org/10.1109/JPHOTOV.2013.2282737} {\bibfield  {journal}
  {\bibinfo  {journal} {Photovoltaics, IEEE Journal of}\ }\textbf {\bibinfo
  {volume} {4}},\ \bibinfo {pages} {96} (\bibinfo {year} {2014})}\BibitemShut
  {NoStop}%
\bibitem [{\citenamefont {Takatsuka}\ \emph {et~al.}(2006)\citenamefont
  {Takatsuka}, \citenamefont {Yamauchi}, \citenamefont {Kawamura},
  \citenamefont {Mashima},\ and\ \citenamefont {Takeuchi}}]{pv-module}%
  \BibitemOpen
  \bibfield  {author} {\bibinfo {author} {\bibfnamefont {H.}~\bibnamefont
  {Takatsuka}}, \bibinfo {author} {\bibfnamefont {Y.}~\bibnamefont {Yamauchi}},
  \bibinfo {author} {\bibfnamefont {K.}~\bibnamefont {Kawamura}}, \bibinfo
  {author} {\bibfnamefont {H.}~\bibnamefont {Mashima}},\ and\ \bibinfo {author}
  {\bibfnamefont {Y.}~\bibnamefont {Takeuchi}},\ }\bibfield  {title} {\bibinfo
  {title} {World's largest amorphous silicon photovoltaic module},\ }\href
  {https://doi.org/http://dx.doi.org/10.1016/j.tsf.2005.08.011} {\bibfield
  {journal} {\bibinfo  {journal} {Thin Solid Films}\ }\textbf {\bibinfo
  {volume} {506}},\ \bibinfo {pages} {13} (\bibinfo {year} {2006})}\BibitemShut
  {NoStop}%
\bibitem [{\citenamefont {Mishima}\ \emph {et~al.}(2011)\citenamefont
  {Mishima}, \citenamefont {Taguchi}, \citenamefont {Sakata},\ and\
  \citenamefont {Maruyama}}]{HIT}%
  \BibitemOpen
  \bibfield  {author} {\bibinfo {author} {\bibfnamefont {T.}~\bibnamefont
  {Mishima}}, \bibinfo {author} {\bibfnamefont {M.}~\bibnamefont {Taguchi}},
  \bibinfo {author} {\bibfnamefont {H.}~\bibnamefont {Sakata}},\ and\ \bibinfo
  {author} {\bibfnamefont {E.}~\bibnamefont {Maruyama}},\ }\bibfield  {title}
  {\bibinfo {title} {Development status of high-efficiency hit solar cells},\
  }\href {http://www.sciencedirect.com/science/article/pii/S0927024810002205}
  {\bibfield  {journal} {\bibinfo  {journal} {Solar Energy Materials and Solar
  Cells}\ }\textbf {\bibinfo {volume} {95}},\ \bibinfo {pages} {18} (\bibinfo
  {year} {2011})}\BibitemShut {NoStop}%
\bibitem [{\citenamefont {Staebler}\ and\ \citenamefont {Wronski}(1977)}]{SW1}%
  \BibitemOpen
  \bibfield  {author} {\bibinfo {author} {\bibfnamefont {D.~L.}\ \bibnamefont
  {Staebler}}\ and\ \bibinfo {author} {\bibfnamefont {C.~R.}\ \bibnamefont
  {Wronski}},\ }\bibfield  {title} {\bibinfo {title} {Reversible conductivity
  changes in discharge‐produced amorphous {Si}},\ }\href
  {https://doi.org/10.1063/1.89674} {\bibfield  {journal} {\bibinfo  {journal}
  {Applied Physics Letters}\ }\textbf {\bibinfo {volume} {31}},\ \bibinfo
  {pages} {292} (\bibinfo {year} {1977})}\BibitemShut {NoStop}%
\bibitem [{\citenamefont {Staebler}\ and\ \citenamefont {Wronski}(1980)}]{SW2}%
  \BibitemOpen
  \bibfield  {author} {\bibinfo {author} {\bibfnamefont {D.~L.}\ \bibnamefont
  {Staebler}}\ and\ \bibinfo {author} {\bibfnamefont {C.~R.}\ \bibnamefont
  {Wronski}},\ }\bibfield  {title} {\bibinfo {title} {Optically induced
  conductivity changes in discharge‐produced hydrogenated amorphous
  silicon},\ }\href {https://doi.org/10.1063/1.328084} {\bibfield  {journal}
  {\bibinfo  {journal} {Journal of Applied Physics}\ }\textbf {\bibinfo
  {volume} {51}},\ \bibinfo {pages} {3262} (\bibinfo {year}
  {1980})}\BibitemShut {NoStop}%
\bibitem [{\citenamefont {Biswas}\ \emph {et~al.}(2007)\citenamefont {Biswas},
  \citenamefont {Atta-Fynn},\ and\ \citenamefont {Drabold}}]{ECMR07}%
  \BibitemOpen
  \bibfield  {author} {\bibinfo {author} {\bibfnamefont {P.}~\bibnamefont
  {Biswas}}, \bibinfo {author} {\bibfnamefont {R.}~\bibnamefont {Atta-Fynn}},\
  and\ \bibinfo {author} {\bibfnamefont {D.~A.}\ \bibnamefont {Drabold}},\
  }\bibfield  {title} {\bibinfo {title} {Experimentally constrained molecular
  relaxation: The case of hydrogenated amorphous silicon},\ }\href
  {https://doi.org/10.1103/PhysRevB.76.125210} {\bibfield  {journal} {\bibinfo
  {journal} {Phys. Rev. B}\ }\textbf {\bibinfo {volume} {76}},\ \bibinfo
  {pages} {125210} (\bibinfo {year} {2007})}\BibitemShut {NoStop}%
\bibitem [{\citenamefont {Biswas}\ \emph {et~al.}(2016)\citenamefont {Biswas},
  \citenamefont {Atta-Fynn},\ and\ \citenamefont {Elliott}}]{Meta2016}%
  \BibitemOpen
  \bibfield  {author} {\bibinfo {author} {\bibfnamefont {P.}~\bibnamefont
  {Biswas}}, \bibinfo {author} {\bibfnamefont {R.}~\bibnamefont {Atta-Fynn}},\
  and\ \bibinfo {author} {\bibfnamefont {S.~R.}\ \bibnamefont {Elliott}},\
  }\bibfield  {title} {\bibinfo {title} {Metadynamical approach to the
  generation of amorphous structures: The case of $a$-{S}i:{H}},\ }\href
  {https://doi.org/10.1103/PhysRevB.93.184202} {\bibfield  {journal} {\bibinfo
  {journal} {Phys. Rev. B}\ }\textbf {\bibinfo {volume} {93}},\ \bibinfo
  {pages} {184202} (\bibinfo {year} {2016})}\BibitemShut {NoStop}%
\bibitem [{\citenamefont {Biswas}\ \emph {et~al.}(2017)\citenamefont {Biswas},
  \citenamefont {Paudel}, \citenamefont {Atta-Fynn}, \citenamefont {Drabold},\
  and\ \citenamefont {Elliott}}]{Biswas2017}%
  \BibitemOpen
  \bibfield  {author} {\bibinfo {author} {\bibfnamefont {P.}~\bibnamefont
  {Biswas}}, \bibinfo {author} {\bibfnamefont {D.}~\bibnamefont {Paudel}},
  \bibinfo {author} {\bibfnamefont {R.}~\bibnamefont {Atta-Fynn}}, \bibinfo
  {author} {\bibfnamefont {D.~A.}\ \bibnamefont {Drabold}},\ and\ \bibinfo
  {author} {\bibfnamefont {S.~R.}\ \bibnamefont {Elliott}},\ }\bibfield
  {title} {\bibinfo {title} {Morphology and number density of voids in
  hydrogenated amorphous silicon: An ab initio study},\ }\href
  {https://doi.org/10.1103/PhysRevApplied.7.024013} {\bibfield  {journal}
  {\bibinfo  {journal} {Phys. Rev. Applied}\ }\textbf {\bibinfo {volume} {7}},\
  \bibinfo {pages} {024013} (\bibinfo {year} {2017})}\BibitemShut {NoStop}%
\bibitem [{\citenamefont {Biswas}\ and\ \citenamefont
  {Timilsina}(2011)}]{Biswas2011}%
  \BibitemOpen
  \bibfield  {author} {\bibinfo {author} {\bibfnamefont {P.}~\bibnamefont
  {Biswas}}\ and\ \bibinfo {author} {\bibfnamefont {R.}~\bibnamefont
  {Timilsina}},\ }\bibfield  {title} {\bibinfo {title} {Vacancies,
  microstructure and the moments of nuclear magnetic resonance: the case of
  hydrogenated amorphous silicon},\ }\href
  {https://doi.org/10.1088/0953-8984/23/6/065801} {\bibfield  {journal}
  {\bibinfo  {journal} {J. Phys.: Cond. Matt.}\ }\textbf {\bibinfo {volume}
  {23}},\ \bibinfo {pages} {065801} (\bibinfo {year} {2011})}\BibitemShut
  {NoStop}%
\bibitem [{\citenamefont {Biswas}\ \emph {et~al.}(2020)\citenamefont {Biswas},
  \citenamefont {Paudel}, \citenamefont {Atta-Fynn},\ and\ \citenamefont
  {Elliott}}]{Biswas2020}%
  \BibitemOpen
  \bibfield  {author} {\bibinfo {author} {\bibfnamefont {P.}~\bibnamefont
  {Biswas}}, \bibinfo {author} {\bibfnamefont {D.}~\bibnamefont {Paudel}},
  \bibinfo {author} {\bibfnamefont {R.}~\bibnamefont {Atta-Fynn}},\ and\
  \bibinfo {author} {\bibfnamefont {S.~R.}\ \bibnamefont {Elliott}},\
  }\bibfield  {title} {\bibinfo {title} {Temperature-induced nanostructural
  evolution of hydrogen-rich voids in amorphous silicon: a first-principles
  study},\ }\href {https://doi.org/10.1039/C9NR08209C} {\bibfield  {journal}
  {\bibinfo  {journal} {Nanoscale}\ }\textbf {\bibinfo {volume} {12}},\
  \bibinfo {pages} {1464} (\bibinfo {year} {2020})}\BibitemShut {NoStop}%
\bibitem [{\citenamefont {Biswas}\ and\ \citenamefont
  {Limbu}(2021)}]{Biswas2021}%
  \BibitemOpen
  \bibfield  {author} {\bibinfo {author} {\bibfnamefont {P.}~\bibnamefont
  {Biswas}}\ and\ \bibinfo {author} {\bibfnamefont {D.}~\bibnamefont {Limbu}},\
  }\bibfield  {title} {\bibinfo {title} {Ab initio hydrogen dynamics and the
  morphology of voids in amorphous silicon},\ }\href
  {https://doi.org/https://doi.org/10.1002/pssb.202170047} {\bibfield
  {journal} {\bibinfo  {journal} {Physica Status Solidi (b)}\ }\textbf
  {\bibinfo {volume} {258}},\ \bibinfo {pages} {2170047} (\bibinfo {year}
  {2021})}\BibitemShut {NoStop}%
\bibitem [{\citenamefont {Min}\ \emph {et~al.}(1992)\citenamefont {Min},
  \citenamefont {Lee}, \citenamefont {Wang}, \citenamefont {Chan},\ and\
  \citenamefont {Ho}}]{Min}%
  \BibitemOpen
  \bibfield  {author} {\bibinfo {author} {\bibfnamefont {B.~J.}\ \bibnamefont
  {Min}}, \bibinfo {author} {\bibfnamefont {Y.~H.}\ \bibnamefont {Lee}},
  \bibinfo {author} {\bibfnamefont {C.~Z.}\ \bibnamefont {Wang}}, \bibinfo
  {author} {\bibfnamefont {C.~T.}\ \bibnamefont {Chan}},\ and\ \bibinfo
  {author} {\bibfnamefont {K.~M.}\ \bibnamefont {Ho}},\ }\bibfield  {title}
  {\bibinfo {title} {Tight-binding model for hydrogen-silicon interactions},\
  }\href {https://doi.org/10.1103/PhysRevB.45.6839} {\bibfield  {journal}
  {\bibinfo  {journal} {Phys. Rev. B}\ }\textbf {\bibinfo {volume} {45}},\
  \bibinfo {pages} {6839} (\bibinfo {year} {1992})}\BibitemShut {NoStop}%
\bibitem [{\citenamefont {Holender}\ \emph {et~al.}(1993)\citenamefont
  {Holender}, \citenamefont {Morgan},\ and\ \citenamefont {Jones}}]{Holender}%
  \BibitemOpen
  \bibfield  {author} {\bibinfo {author} {\bibfnamefont {J.~M.}\ \bibnamefont
  {Holender}}, \bibinfo {author} {\bibfnamefont {G.~J.}\ \bibnamefont
  {Morgan}},\ and\ \bibinfo {author} {\bibfnamefont {R.}~\bibnamefont
  {Jones}},\ }\bibfield  {title} {\bibinfo {title} {Model of hydrogenated
  amorphous silicon and its electronic structure},\ }\href
  {https://doi.org/10.1103/PhysRevB.47.3991} {\bibfield  {journal} {\bibinfo
  {journal} {Phys. Rev. B}\ }\textbf {\bibinfo {volume} {47}},\ \bibinfo
  {pages} {3991} (\bibinfo {year} {1993})}\BibitemShut {NoStop}%
\bibitem [{\citenamefont {Tuttle}\ and\ \citenamefont {Adams}(1998)}]{Tuttle}%
  \BibitemOpen
  \bibfield  {author} {\bibinfo {author} {\bibfnamefont {B.}~\bibnamefont
  {Tuttle}}\ and\ \bibinfo {author} {\bibfnamefont {J.~B.}\ \bibnamefont
  {Adams}},\ }\bibfield  {title} {\bibinfo {title} {Energetics of hydrogen in
  amorphous silicon: An \textit{ab initio} study},\ }\href@noop {} {\bibfield
  {journal} {\bibinfo  {journal} {Phys. Rev. B}\ }\textbf {\bibinfo {volume}
  {57}},\ \bibinfo {pages} {12859} (\bibinfo {year} {1998})}\BibitemShut
  {NoStop}%
\bibitem [{\citenamefont {Buda}\ \emph {et~al.}(1991)\citenamefont {Buda},
  \citenamefont {Chiarotti}, \citenamefont {Car},\ and\ \citenamefont
  {Parrinello}}]{buda}%
  \BibitemOpen
  \bibfield  {author} {\bibinfo {author} {\bibfnamefont {F.}~\bibnamefont
  {Buda}}, \bibinfo {author} {\bibfnamefont {G.~L.}\ \bibnamefont {Chiarotti}},
  \bibinfo {author} {\bibfnamefont {R.}~\bibnamefont {Car}},\ and\ \bibinfo
  {author} {\bibfnamefont {M.}~\bibnamefont {Parrinello}},\ }\bibfield  {title}
  {\bibinfo {title} {Structure of hydrogenated amorphous silicon from
  \textit{ab} \textit{initio} molecular dynamics},\ }\href
  {https://doi.org/10.1103/PhysRevB.44.5908} {\bibfield  {journal} {\bibinfo
  {journal} {Phys. Rev. B}\ }\textbf {\bibinfo {volume} {44}},\ \bibinfo
  {pages} {5908} (\bibinfo {year} {1991})}\BibitemShut {NoStop}%
\bibitem [{\citenamefont {Jarolimek}\ \emph {et~al.}(2009)\citenamefont
  {Jarolimek}, \citenamefont {de~Groot}, \citenamefont {de~Wijs},\ and\
  \citenamefont {Zeman}}]{Jarolimek}%
  \BibitemOpen
  \bibfield  {author} {\bibinfo {author} {\bibfnamefont {K.}~\bibnamefont
  {Jarolimek}}, \bibinfo {author} {\bibfnamefont {R.~A.}\ \bibnamefont
  {de~Groot}}, \bibinfo {author} {\bibfnamefont {G.~A.}\ \bibnamefont
  {de~Wijs}},\ and\ \bibinfo {author} {\bibfnamefont {M.}~\bibnamefont
  {Zeman}},\ }\bibfield  {title} {\bibinfo {title} {First-principles study of
  hydrogenated amorphous silicon},\ }\href
  {https://doi.org/10.1103/PhysRevB.79.155206} {\bibfield  {journal} {\bibinfo
  {journal} {Phys. Rev. B}\ }\textbf {\bibinfo {volume} {79}},\ \bibinfo
  {pages} {155206} (\bibinfo {year} {2009})}\BibitemShut {NoStop}%
\bibitem [{\citenamefont {Drabold}\ \emph {et~al.}(1991)\citenamefont
  {Drabold}, \citenamefont {Fedders}, \citenamefont {Klemm},\ and\
  \citenamefont {Sankey}}]{drabold1}%
  \BibitemOpen
  \bibfield  {author} {\bibinfo {author} {\bibfnamefont {D.~A.}\ \bibnamefont
  {Drabold}}, \bibinfo {author} {\bibfnamefont {P.~A.}\ \bibnamefont
  {Fedders}}, \bibinfo {author} {\bibfnamefont {S.}~\bibnamefont {Klemm}},\
  and\ \bibinfo {author} {\bibfnamefont {O.~F.}\ \bibnamefont {Sankey}},\
  }\bibfield  {title} {\bibinfo {title} {Finite-temperature properties of
  amorphous silicon},\ }\href {https://doi.org/10.1103/PhysRevLett.67.2179}
  {\bibfield  {journal} {\bibinfo  {journal} {Phys. Rev. Lett.}\ }\textbf
  {\bibinfo {volume} {67}},\ \bibinfo {pages} {2179} (\bibinfo {year}
  {1991})}\BibitemShut {NoStop}%
\bibitem [{\citenamefont {Klein}\ \emph {et~al.}(1999)\citenamefont {Klein},
  \citenamefont {Urbassek},\ and\ \citenamefont {Frauenheim}}]{klein}%
  \BibitemOpen
  \bibfield  {author} {\bibinfo {author} {\bibfnamefont {P.}~\bibnamefont
  {Klein}}, \bibinfo {author} {\bibfnamefont {H.~M.}\ \bibnamefont
  {Urbassek}},\ and\ \bibinfo {author} {\bibfnamefont {T.}~\bibnamefont
  {Frauenheim}},\ }\bibfield  {title} {\bibinfo {title} {Tight-binding
  molecular-dynamics study of $a$-{Si}:{H}: Preparation, structure, and
  dynamics},\ }\href {https://doi.org/10.1103/PhysRevB.60.5478} {\bibfield
  {journal} {\bibinfo  {journal} {Phys. Rev. B}\ }\textbf {\bibinfo {volume}
  {60}},\ \bibinfo {pages} {5478} (\bibinfo {year} {1999})}\BibitemShut
  {NoStop}%
\bibitem [{\citenamefont {Deringer}\ \emph {et~al.}(2018)\citenamefont
  {Deringer}, \citenamefont {Bernstein}, \citenamefont {Bart\'ok},
  \citenamefont {Cliffe}, \citenamefont {Kerber}, \citenamefont {Marbella},
  \citenamefont {Grey}, \citenamefont {Elliott},\ and\ \citenamefont
  {Cs\'anyi}}]{Deringer2018}%
  \BibitemOpen
  \bibfield  {author} {\bibinfo {author} {\bibfnamefont {V.~L.}\ \bibnamefont
  {Deringer}}, \bibinfo {author} {\bibfnamefont {N.}~\bibnamefont {Bernstein}},
  \bibinfo {author} {\bibfnamefont {A.~P.}\ \bibnamefont {Bart\'ok}}, \bibinfo
  {author} {\bibfnamefont {M.~J.}\ \bibnamefont {Cliffe}}, \bibinfo {author}
  {\bibfnamefont {R.~N.}\ \bibnamefont {Kerber}}, \bibinfo {author}
  {\bibfnamefont {L.~E.}\ \bibnamefont {Marbella}}, \bibinfo {author}
  {\bibfnamefont {C.~P.}\ \bibnamefont {Grey}}, \bibinfo {author}
  {\bibfnamefont {S.~R.}\ \bibnamefont {Elliott}},\ and\ \bibinfo {author}
  {\bibfnamefont {G.}~\bibnamefont {Cs\'anyi}},\ }\bibfield  {title} {\bibinfo
  {title} {Realistic atomistic structure of amorphous silicon from
  machine-learning-driven molecular dynamics},\ }\href
  {https://doi.org/10.1021/acs.jpclett.8b00902} {\bibfield  {journal} {\bibinfo
   {journal} {J. Phys. Chem. Lett.}\ }\textbf {\bibinfo {volume} {9}},\
  \bibinfo {pages} {2879} (\bibinfo {year} {2018})}\BibitemShut {NoStop}%
\bibitem [{\citenamefont {Biswas}\ \emph {et~al.}(2004)\citenamefont {Biswas},
  \citenamefont {Atta-Fynn},\ and\ \citenamefont {Drabold}}]{RMC04}%
  \BibitemOpen
  \bibfield  {author} {\bibinfo {author} {\bibfnamefont {P.}~\bibnamefont
  {Biswas}}, \bibinfo {author} {\bibfnamefont {R.}~\bibnamefont {Atta-Fynn}},\
  and\ \bibinfo {author} {\bibfnamefont {D.~A.}\ \bibnamefont {Drabold}},\
  }\bibfield  {title} {\bibinfo {title} {Reverse {M}onte {C}arlo modeling of
  amorphous silicon},\ }\href {https://doi.org/10.1103/PhysRevB.69.195207}
  {\bibfield  {journal} {\bibinfo  {journal} {Phys. Rev. B}\ }\textbf {\bibinfo
  {volume} {69}},\ \bibinfo {pages} {195207} (\bibinfo {year}
  {2004})}\BibitemShut {NoStop}%
\bibitem [{\citenamefont {Walters}\ and\ \citenamefont
  {Newport}(1996)}]{RMC96}%
  \BibitemOpen
  \bibfield  {author} {\bibinfo {author} {\bibfnamefont {J.~K.}\ \bibnamefont
  {Walters}}\ and\ \bibinfo {author} {\bibfnamefont {R.~J.}\ \bibnamefont
  {Newport}},\ }\bibfield  {title} {\bibinfo {title} {Reverse {Monte} {C}arlo
  modeling of amorphous germanium},\ }\href
  {https://doi.org/10.1103/PhysRevB.53.2405} {\bibfield  {journal} {\bibinfo
  {journal} {Phys. Rev. B}\ }\textbf {\bibinfo {volume} {53}},\ \bibinfo
  {pages} {2405} (\bibinfo {year} {1996})}\BibitemShut {NoStop}%
\bibitem [{\citenamefont {Limbu}\ \emph {et~al.}(2020)\citenamefont {Limbu},
  \citenamefont {Elliott}, \citenamefont {Atta-Fynn},\ and\ \citenamefont
  {Biswas}}]{CMC19}%
  \BibitemOpen
  \bibfield  {author} {\bibinfo {author} {\bibfnamefont {D.~K.}\ \bibnamefont
  {Limbu}}, \bibinfo {author} {\bibfnamefont {S.~R.}\ \bibnamefont {Elliott}},
  \bibinfo {author} {\bibfnamefont {R.}~\bibnamefont {Atta-Fynn}},\ and\
  \bibinfo {author} {\bibfnamefont {P.}~\bibnamefont {Biswas}},\ }\bibfield
  {title} {\bibinfo {title} {Disorder by design: A data-driven approach to
  amorphous semiconductors without total-energy functionals},\ }\href
  {https://doi.org/10.1038/s41598-020-64327-3} {\bibfield  {journal} {\bibinfo
  {journal} {Scientific Reports}\ }\textbf {\bibinfo {volume} {10}},\ \bibinfo
  {pages} {7742} (\bibinfo {year} {2020})}\BibitemShut {NoStop}%
\bibitem [{\citenamefont {Biswas}\ \emph {et~al.}(2005)\citenamefont {Biswas},
  \citenamefont {Tafen},\ and\ \citenamefont {Drabold}}]{ECMR05}%
  \BibitemOpen
  \bibfield  {author} {\bibinfo {author} {\bibfnamefont {P.}~\bibnamefont
  {Biswas}}, \bibinfo {author} {\bibfnamefont {D.~N.}\ \bibnamefont {Tafen}},\
  and\ \bibinfo {author} {\bibfnamefont {D.~A.}\ \bibnamefont {Drabold}},\
  }\bibfield  {title} {\bibinfo {title} {Experimentally constrained molecular
  relaxation: The case of glassy {Ge}{Se}$_2$},\ }\href
  {https://doi.org/10.1103/PhysRevB.71.054204} {\bibfield  {journal} {\bibinfo
  {journal} {Phys. Rev. B}\ }\textbf {\bibinfo {volume} {71}},\ \bibinfo
  {pages} {054204} (\bibinfo {year} {2005})}\BibitemShut {NoStop}%
\bibitem [{\citenamefont {Pandey}\ \emph {et~al.}(2016)\citenamefont {Pandey},
  \citenamefont {Biswas},\ and\ \citenamefont {Drabold}}]{FEAR-SR}%
  \BibitemOpen
  \bibfield  {author} {\bibinfo {author} {\bibfnamefont {A.}~\bibnamefont
  {Pandey}}, \bibinfo {author} {\bibfnamefont {P.}~\bibnamefont {Biswas}},\
  and\ \bibinfo {author} {\bibfnamefont {D.~A.}\ \bibnamefont {Drabold}},\
  }\bibfield  {title} {\bibinfo {title} {Inversion of diffraction data for
  amorphous materials},\ }\href {https://doi.org/10.1038/srep33731} {\bibfield
  {journal} {\bibinfo  {journal} {Sci. Rep.}\ }\textbf {\bibinfo {volume}
  {6}},\ \bibinfo {pages} {33731} (\bibinfo {year} {2016})}\BibitemShut
  {NoStop}%
\bibitem [{\citenamefont {Pandey}\ \emph {et~al.}(2015)\citenamefont {Pandey},
  \citenamefont {Biswas},\ and\ \citenamefont {Drabold}}]{FEAR-PRB}%
  \BibitemOpen
  \bibfield  {author} {\bibinfo {author} {\bibfnamefont {A.}~\bibnamefont
  {Pandey}}, \bibinfo {author} {\bibfnamefont {P.}~\bibnamefont {Biswas}},\
  and\ \bibinfo {author} {\bibfnamefont {D.~A.}\ \bibnamefont {Drabold}},\
  }\bibfield  {title} {\bibinfo {title} {Force-enhanced atomic refinement:
  Structural modeling with interatomic forces in a reverse monte carlo approach
  applied to amorphous {Si} and {Si}{O}$_2$},\ }\href
  {https://doi.org/10.1103/PhysRevB.92.155205} {\bibfield  {journal} {\bibinfo
  {journal} {Phys. Rev. B}\ }\textbf {\bibinfo {volume} {92}},\ \bibinfo
  {pages} {155205} (\bibinfo {year} {2015})}\BibitemShut {NoStop}%
\bibitem [{\citenamefont {Limbu}\ \emph {et~al.}(2018)\citenamefont {Limbu},
  \citenamefont {Atta-Fynn}, \citenamefont {Drabold}, \citenamefont {Elliott},\
  and\ \citenamefont {Biswas}}]{INDIA}%
  \BibitemOpen
  \bibfield  {author} {\bibinfo {author} {\bibfnamefont {D.~K.}\ \bibnamefont
  {Limbu}}, \bibinfo {author} {\bibfnamefont {R.}~\bibnamefont {Atta-Fynn}},
  \bibinfo {author} {\bibfnamefont {D.~A.}\ \bibnamefont {Drabold}}, \bibinfo
  {author} {\bibfnamefont {S.~R.}\ \bibnamefont {Elliott}},\ and\ \bibinfo
  {author} {\bibfnamefont {P.}~\bibnamefont {Biswas}},\ }\bibfield  {title}
  {\bibinfo {title} {Information-driven inverse approach to disordered solids:
  Applications to amorphous silicon},\ }\href
  {https://doi.org/10.1103/PhysRevMaterials.2.115602} {\bibfield  {journal}
  {\bibinfo  {journal} {Phys. Rev. Materials}\ }\textbf {\bibinfo {volume}
  {2}},\ \bibinfo {pages} {115602} (\bibinfo {year} {2018})}\BibitemShut
  {NoStop}%
\bibitem [{\citenamefont {Biswas}\ and\ \citenamefont {Elliott}(2015)}]{srepb}%
  \BibitemOpen
  \bibfield  {author} {\bibinfo {author} {\bibfnamefont {P.}~\bibnamefont
  {Biswas}}\ and\ \bibinfo {author} {\bibfnamefont {S.~R.}\ \bibnamefont
  {Elliott}},\ }\bibfield  {title} {\bibinfo {title} {Nanoscale structure of
  microvoids in $a$-{Si}:{H}: {A} first-principles study},\ }\href
  {http://stacks.iop.org/0953-8984/27/i=43/a=435201} {\bibfield  {journal}
  {\bibinfo  {journal} {J. Phys.: Cond. Matt.}\ }\textbf {\bibinfo {volume}
  {27}},\ \bibinfo {pages} {435201} (\bibinfo {year} {2015})}\BibitemShut
  {NoStop}%
\bibitem [{\citenamefont {Carlos}\ and\ \citenamefont {Taylor}(1982)}]{Taylor}%
  \BibitemOpen
  \bibfield  {author} {\bibinfo {author} {\bibfnamefont {W.~E.}\ \bibnamefont
  {Carlos}}\ and\ \bibinfo {author} {\bibfnamefont {P.~C.}\ \bibnamefont
  {Taylor}},\ }\bibfield  {title} {\bibinfo {title} {$^{1}\mathrm{H}$ {N}{M}{R}
  in $a$-{S}i},\ }\href {https://doi.org/10.1103/PhysRevB.26.3605} {\bibfield
  {journal} {\bibinfo  {journal} {Phys. Rev. B}\ }\textbf {\bibinfo {volume}
  {26}},\ \bibinfo {pages} {3605} (\bibinfo {year} {1982})}\BibitemShut
  {NoStop}%
\bibitem [{\citenamefont {Reimer}\ \emph {et~al.}(1980)\citenamefont {Reimer},
  \citenamefont {Vaughan},\ and\ \citenamefont {Knights}}]{nmr1}%
  \BibitemOpen
  \bibfield  {author} {\bibinfo {author} {\bibfnamefont {J.~A.}\ \bibnamefont
  {Reimer}}, \bibinfo {author} {\bibfnamefont {R.~W.}\ \bibnamefont
  {Vaughan}},\ and\ \bibinfo {author} {\bibfnamefont {J.~C.}\ \bibnamefont
  {Knights}},\ }\bibfield  {title} {\bibinfo {title} {Proton magnetic resonance
  spectra of plasma-deposited amorphous si: H films},\ }\href
  {https://doi.org/10.1103/PhysRevLett.44.193} {\bibfield  {journal} {\bibinfo
  {journal} {Phys. Rev. Lett.}\ }\textbf {\bibinfo {volume} {44}},\ \bibinfo
  {pages} {193} (\bibinfo {year} {1980})}\BibitemShut {NoStop}%
\bibitem [{\citenamefont {Leopold}\ \emph {et~al.}(1982)\citenamefont
  {Leopold}, \citenamefont {Boyce}, \citenamefont {Fedders},\ and\
  \citenamefont {Norberg}}]{nmr2}%
  \BibitemOpen
  \bibfield  {author} {\bibinfo {author} {\bibfnamefont {D.~J.}\ \bibnamefont
  {Leopold}}, \bibinfo {author} {\bibfnamefont {J.~B.}\ \bibnamefont {Boyce}},
  \bibinfo {author} {\bibfnamefont {P.~A.}\ \bibnamefont {Fedders}},\ and\
  \bibinfo {author} {\bibfnamefont {R.~E.}\ \bibnamefont {Norberg}},\
  }\bibfield  {title} {\bibinfo {title} {Deuteron and proton magnetic resonance
  in $a$-{S}i:({D},{H})},\ }\href {https://doi.org/10.1103/PhysRevB.26.6053}
  {\bibfield  {journal} {\bibinfo  {journal} {Phys. Rev. B}\ }\textbf {\bibinfo
  {volume} {26}},\ \bibinfo {pages} {6053} (\bibinfo {year}
  {1982})}\BibitemShut {NoStop}%
\bibitem [{\citenamefont {Baum}\ \emph {et~al.}(1986)\citenamefont {Baum},
  \citenamefont {Gleason}, \citenamefont {Pines}, \citenamefont {Garroway},\
  and\ \citenamefont {Reimer}}]{Baum}%
  \BibitemOpen
  \bibfield  {author} {\bibinfo {author} {\bibfnamefont {J.}~\bibnamefont
  {Baum}}, \bibinfo {author} {\bibfnamefont {K.~K.}\ \bibnamefont {Gleason}},
  \bibinfo {author} {\bibfnamefont {A.}~\bibnamefont {Pines}}, \bibinfo
  {author} {\bibfnamefont {A.~N.}\ \bibnamefont {Garroway}},\ and\ \bibinfo
  {author} {\bibfnamefont {J.~A.}\ \bibnamefont {Reimer}},\ }\bibfield  {title}
  {\bibinfo {title} {Multiple-quantum {N}{M}{R} study of clustering in
  hydrogenated amorphous silicon},\ }\href
  {https://doi.org/10.1103/PhysRevLett.56.1377} {\bibfield  {journal} {\bibinfo
   {journal} {Phys. Rev. Lett.}\ }\textbf {\bibinfo {volume} {56}},\ \bibinfo
  {pages} {1377} (\bibinfo {year} {1986})}\BibitemShut {NoStop}%
\bibitem [{\citenamefont {Chabal}\ and\ \citenamefont
  {Patel}(1984)}]{Chabal1984}%
  \BibitemOpen
  \bibfield  {author} {\bibinfo {author} {\bibfnamefont {Y.~J.}\ \bibnamefont
  {Chabal}}\ and\ \bibinfo {author} {\bibfnamefont {C.~K.~N.}\ \bibnamefont
  {Patel}},\ }\bibfield  {title} {\bibinfo {title} {Infrared {Absorption} in
  $a$-{Si}:{H}: {First} {Observation} of {Gaseous} {Molecular} {H}$_2$ and
  {Si}-{H} {Overtone}},\ }\href {https://doi.org/10.1103/PhysRevLett.53.210}
  {\bibfield  {journal} {\bibinfo  {journal} {Phys. Rev. Lett.}\ }\textbf
  {\bibinfo {volume} {53}},\ \bibinfo {pages} {210} (\bibinfo {year}
  {1984})}\BibitemShut {NoStop}%
\bibitem [{\citenamefont {Scharff}\ and\ \citenamefont
  {McGrane}(2007)}]{Scharff}%
  \BibitemOpen
  \bibfield  {author} {\bibinfo {author} {\bibfnamefont {R.~J.}\ \bibnamefont
  {Scharff}}\ and\ \bibinfo {author} {\bibfnamefont {S.~D.}\ \bibnamefont
  {McGrane}},\ }\bibfield  {title} {\bibinfo {title} {Si-{H} bond dynamics in
  hydrogenated amorphous silicon},\ }\href
  {https://doi.org/10.1103/PhysRevB.76.054301} {\bibfield  {journal} {\bibinfo
  {journal} {Phys. Rev. B}\ }\textbf {\bibinfo {volume} {76}},\ \bibinfo
  {pages} {054301} (\bibinfo {year} {2007})}\BibitemShut {NoStop}%
\bibitem [{\citenamefont {Kageyama}\ \emph {et~al.}(2011)\citenamefont
  {Kageyama}, \citenamefont {Akagawa},\ and\ \citenamefont
  {Fujiwara}}]{Kageyama}%
  \BibitemOpen
  \bibfield  {author} {\bibinfo {author} {\bibfnamefont {S.}~\bibnamefont
  {Kageyama}}, \bibinfo {author} {\bibfnamefont {M.}~\bibnamefont {Akagawa}},\
  and\ \bibinfo {author} {\bibfnamefont {H.}~\bibnamefont {Fujiwara}},\
  }\bibfield  {title} {\bibinfo {title} {Dielectric function of $a$-{Si}:{H}
  based on local network structures},\ }\href
  {https://doi.org/10.1103/PhysRevB.83.195205} {\bibfield  {journal} {\bibinfo
  {journal} {Phys. Rev. B}\ }\textbf {\bibinfo {volume} {83}},\ \bibinfo
  {pages} {195205} (\bibinfo {year} {2011})}\BibitemShut {NoStop}%
\bibitem [{\citenamefont {Kamitakahara}\ \emph {et~al.}(1984)\citenamefont
  {Kamitakahara}, \citenamefont {Shanks}, \citenamefont {McClelland},
  \citenamefont {Buchenau}, \citenamefont {Gompf},\ and\ \citenamefont
  {Pintschovius}}]{Kamita}%
  \BibitemOpen
  \bibfield  {author} {\bibinfo {author} {\bibfnamefont {W.~A.}\ \bibnamefont
  {Kamitakahara}}, \bibinfo {author} {\bibfnamefont {H.~R.}\ \bibnamefont
  {Shanks}}, \bibinfo {author} {\bibfnamefont {J.~F.}\ \bibnamefont
  {McClelland}}, \bibinfo {author} {\bibfnamefont {U.}~\bibnamefont
  {Buchenau}}, \bibinfo {author} {\bibfnamefont {F.}~\bibnamefont {Gompf}},\
  and\ \bibinfo {author} {\bibfnamefont {L.}~\bibnamefont {Pintschovius}},\
  }\bibfield  {title} {\bibinfo {title} {Measurement of phonon densities of
  states for pure and hydrogenated amorphous silicon},\ }\href
  {https://doi.org/10.1103/PhysRevLett.52.644} {\bibfield  {journal} {\bibinfo
  {journal} {Phys. Rev. Lett.}\ }\textbf {\bibinfo {volume} {52}},\ \bibinfo
  {pages} {644} (\bibinfo {year} {1984})}\BibitemShut {NoStop}%
\bibitem [{\citenamefont {Laio}\ and\ \citenamefont {Parrinello}(2002)}]{Laio}%
  \BibitemOpen
  \bibfield  {author} {\bibinfo {author} {\bibfnamefont {A.}~\bibnamefont
  {Laio}}\ and\ \bibinfo {author} {\bibfnamefont {M.}~\bibnamefont
  {Parrinello}},\ }\href {https://doi.org/10.1073/pnas.20242739} {\bibfield
  {journal} {\bibinfo  {journal} {Proc. Natl. Acad. Sci. U.S.A.}\ }\textbf
  {\bibinfo {volume} {99}},\ \bibinfo {pages} {12562} (\bibinfo {year}
  {2002})}\BibitemShut {NoStop}%
\bibitem [{\citenamefont {Lucy}(1977)}]{Lucy_1977}%
  \BibitemOpen
  \bibfield  {author} {\bibinfo {author} {\bibfnamefont {L.~B.}\ \bibnamefont
  {Lucy}},\ }\bibfield  {title} {\bibinfo {title} {A numerical approach to the
  testing of the fission hypothesis},\ }\href {https://doi.org/10.1086/112164}
  {\bibfield  {journal} {\bibinfo  {journal} {The Astronomical Journal}\
  }\textbf {\bibinfo {volume} {82}},\ \bibinfo {pages} {1013} (\bibinfo {year}
  {1977})}\BibitemShut {NoStop}%
\bibitem [{nor()}]{normalization}%
  \BibitemOpen
  \href@noop {} {}\bibinfo {note} {If $\mathcal{L}$ is normalized (that is
  $\int\mathcal{L}dx=1$ in 1D, $\iint\mathcal{L}dxdy=1$ in 2D, and
  $\iiint\mathcal{L}dxdydz=1$ in 3D), then $A=5/(4w)$ in 1D, $A=5/(\pi w^2)$ in
  2D, and $A=105/(16\pi w^3)$ in 3D.}\BibitemShut {Stop}%
\bibitem [{luc()}]{lucy}%
  \BibitemOpen
  \href@noop {} {}\bibinfo {note} {The relation between $\sigma$ and $w$ is
  based on the condition that $\mathcal{L}$ and $g$ both have the same full
  width at half maximum (FWHM).}\BibitemShut {Stop}%
\bibitem [{\citenamefont {Atta-Fynn}\ and\ \citenamefont
  {Biswas}(2018)}]{raf_asi}%
  \BibitemOpen
  \bibfield  {author} {\bibinfo {author} {\bibfnamefont {R.}~\bibnamefont
  {Atta-Fynn}}\ and\ \bibinfo {author} {\bibfnamefont {P.}~\bibnamefont
  {Biswas}},\ }\bibfield  {title} {\bibinfo {title} {Nearly defect-free
  dynamical models of disordered solids: The case of amorphous silicon},\
  }\href {https://doi.org/10.1063/1.5021813} {\bibfield  {journal} {\bibinfo
  {journal} {J. Chem. Phys.}\ }\textbf {\bibinfo {volume} {148}},\ \bibinfo
  {pages} {204503} (\bibinfo {year} {2018})}\BibitemShut {NoStop}%
\bibitem [{\citenamefont {Stillinger}\ and\ \citenamefont
  {Weber}(1985)}]{SW-silicon}%
  \BibitemOpen
  \bibfield  {author} {\bibinfo {author} {\bibfnamefont {F.~H.}\ \bibnamefont
  {Stillinger}}\ and\ \bibinfo {author} {\bibfnamefont {T.~A.}\ \bibnamefont
  {Weber}},\ }\bibfield  {title} {\bibinfo {title} {Computer simulation of
  local order in condensed phases of silicon},\ }\href
  {https://doi.org/10.1103/PhysRevB.31.5262} {\bibfield  {journal} {\bibinfo
  {journal} {Phys. Rev. B}\ }\textbf {\bibinfo {volume} {31}},\ \bibinfo
  {pages} {5262} (\bibinfo {year} {1985})}\BibitemShut {NoStop}%
\bibitem [{\citenamefont {Vink}\ \emph {et~al.}(2001)\citenamefont {Vink},
  \citenamefont {Barkema}, \citenamefont {van~der Weg},\ and\ \citenamefont
  {Mousseau}}]{SW-Vink}%
  \BibitemOpen
  \bibfield  {author} {\bibinfo {author} {\bibfnamefont {R.~L.~C.}\
  \bibnamefont {Vink}}, \bibinfo {author} {\bibfnamefont {G.~T.}\ \bibnamefont
  {Barkema}}, \bibinfo {author} {\bibfnamefont {W.~F.}\ \bibnamefont {van~der
  Weg}},\ and\ \bibinfo {author} {\bibfnamefont {N.}~\bibnamefont {Mousseau}},\
  }\bibfield  {title} {\bibinfo {title} {Fitting the stillinge-weber potential
  to amorphous silicon},\ }\href
  {https://doi.org/http://dx.doi.org/10.1016/S0022-3093(01)00342-8} {\bibfield
  {journal} {\bibinfo  {journal} {Journal of Non-Crystalline Solids}\ }\textbf
  {\bibinfo {volume} {282}},\ \bibinfo {pages} {248} (\bibinfo {year}
  {2001})}\BibitemShut {NoStop}%
\bibitem [{MTD()}]{MTD}%
  \BibitemOpen
  \bibinfo {note} {Since metadynamics is a nonequilibrium simulation technique,
  one cannot define simulation temperatures as in conventional MD using the
  classical equipartition theorem. It is more appropriate to regard the
  temperature as a simulation parameter that determines the degree of
  ruggedness of a free-energy landscape defined by a given set of collective
  variables. A high temperature corresponds to a less rugged landscape, and
  vice versa.}\BibitemShut {Stop}%
\bibitem [{\citenamefont {Smets}\ and\ \citenamefont {van~de
  Sanden}(2007)}]{Smets}%
  \BibitemOpen
  \bibfield  {author} {\bibinfo {author} {\bibfnamefont {A.~H.~M.}\
  \bibnamefont {Smets}}\ and\ \bibinfo {author} {\bibfnamefont {M.~C.~M.}\
  \bibnamefont {van~de Sanden}},\ }\bibfield  {title} {\bibinfo {title}
  {Relation of the {Si}-{H} stretching frequency to the nanostructural {Si}{H}
  bulk environment},\ }\href {https://doi.org/10.1103/PhysRevB.76.073202}
  {\bibfield  {journal} {\bibinfo  {journal} {Phys. Rev. B}\ }\textbf {\bibinfo
  {volume} {76}},\ \bibinfo {pages} {073202} (\bibinfo {year}
  {2007})}\BibitemShut {NoStop}%
\bibitem [{\citenamefont {Soler}\ \emph {et~al.}(2002)\citenamefont {Soler},
  \citenamefont {Artacho}, \citenamefont {Gale}, \citenamefont {Garci\'a},
  \citenamefont {Junquera}, \citenamefont {Ordej\'on},\ and\ \citenamefont
  {S\'anchez-Portal}}]{siesta}%
  \BibitemOpen
  \bibfield  {author} {\bibinfo {author} {\bibfnamefont {J.~M.}\ \bibnamefont
  {Soler}}, \bibinfo {author} {\bibfnamefont {E.}~\bibnamefont {Artacho}},
  \bibinfo {author} {\bibfnamefont {J.~D.}\ \bibnamefont {Gale}}, \bibinfo
  {author} {\bibfnamefont {A.}~\bibnamefont {Garci\'a}}, \bibinfo {author}
  {\bibfnamefont {J.}~\bibnamefont {Junquera}}, \bibinfo {author}
  {\bibfnamefont {P.}~\bibnamefont {Ordej\'on}},\ and\ \bibinfo {author}
  {\bibfnamefont {D.}~\bibnamefont {S\'anchez-Portal}},\ }\bibfield  {title}
  {\bibinfo {title} {The siesta method for ab initio order- n materials
  simulation},\ }\href {http://stacks.iop.org/0953-8984/14/i=11/a=302}
  {\bibfield  {journal} {\bibinfo  {journal} {J. Phys. Condens. Matter}\
  }\textbf {\bibinfo {volume} {14}},\ \bibinfo {pages} {2745} (\bibinfo {year}
  {2002})}\BibitemShut {NoStop}%
\bibitem [{\citenamefont {Harris}(1985)}]{harris}%
  \BibitemOpen
  \bibfield  {author} {\bibinfo {author} {\bibfnamefont {J.}~\bibnamefont
  {Harris}},\ }\bibfield  {title} {\bibinfo {title} {Simplified method for
  calculating the energy of weakly interacting fragments},\ }\href
  {https://doi.org/10.1103/PhysRevB.31.1770} {\bibfield  {journal} {\bibinfo
  {journal} {Phys. Rev. B}\ }\textbf {\bibinfo {volume} {31}},\ \bibinfo
  {pages} {1770} (\bibinfo {year} {1985})}\BibitemShut {NoStop}%
\bibitem [{\citenamefont {Atta-Fynn}\ \emph {et~al.}(2004)\citenamefont
  {Atta-Fynn}, \citenamefont {Biswas}, \citenamefont {Ordej\'on},\ and\
  \citenamefont {Drabold}}]{Ray-PRB}%
  \BibitemOpen
  \bibfield  {author} {\bibinfo {author} {\bibfnamefont {R.}~\bibnamefont
  {Atta-Fynn}}, \bibinfo {author} {\bibfnamefont {P.}~\bibnamefont {Biswas}},
  \bibinfo {author} {\bibfnamefont {P.}~\bibnamefont {Ordej\'on}},\ and\
  \bibinfo {author} {\bibfnamefont {D.~A.}\ \bibnamefont {Drabold}},\
  }\bibfield  {title} {\bibinfo {title} {Systematic study of electron
  localization in an amorphous semiconductor},\ }\href
  {https://doi.org/10.1103/PhysRevB.69.085207} {\bibfield  {journal} {\bibinfo
  {journal} {Phys. Rev. B}\ }\textbf {\bibinfo {volume} {69}},\ \bibinfo
  {pages} {085207} (\bibinfo {year} {2004})}\BibitemShut {NoStop}%
\bibitem [{\citenamefont {Ceperley}\ and\ \citenamefont
  {Alder}(1980)}]{Ceperley1980}%
  \BibitemOpen
  \bibfield  {author} {\bibinfo {author} {\bibfnamefont {D.~M.}\ \bibnamefont
  {Ceperley}}\ and\ \bibinfo {author} {\bibfnamefont {B.~J.}\ \bibnamefont
  {Alder}},\ }\bibfield  {title} {\bibinfo {title} {Ground state of the
  electron gas by a stochastic method},\ }\href
  {https://doi.org/10.1103/PhysRevLett.45.566} {\bibfield  {journal} {\bibinfo
  {journal} {Phys. Rev. Lett.}\ }\textbf {\bibinfo {volume} {45}},\ \bibinfo
  {pages} {566} (\bibinfo {year} {1980})}\BibitemShut {NoStop}%
\bibitem [{\citenamefont {Troullier}\ and\ \citenamefont {Martins}(1991)}]{tm}%
  \BibitemOpen
  \bibfield  {author} {\bibinfo {author} {\bibfnamefont {N.}~\bibnamefont
  {Troullier}}\ and\ \bibinfo {author} {\bibfnamefont {J.~L.}\ \bibnamefont
  {Martins}},\ }\bibfield  {title} {\bibinfo {title} {Efficient
  pseudopotentials for plane-wave calculations},\ }\href
  {https://doi.org/10.1103/PhysRevB.43.1993} {\bibfield  {journal} {\bibinfo
  {journal} {Phys. Rev. B}\ }\textbf {\bibinfo {volume} {43}},\ \bibinfo
  {pages} {1993} (\bibinfo {year} {1991})}\BibitemShut {NoStop}%
\bibitem [{\citenamefont {Ouwens}\ and\ \citenamefont {Schropp}(1996)}]{RS}%
  \BibitemOpen
  \bibfield  {author} {\bibinfo {author} {\bibfnamefont {J.~D.}\ \bibnamefont
  {Ouwens}}\ and\ \bibinfo {author} {\bibfnamefont {R.~E.~I.}\ \bibnamefont
  {Schropp}},\ }\bibfield  {title} {\bibinfo {title} {Hydrogen microstructure
  in hydrogenated amorphous silicon},\ }\href
  {https://doi.org/10.1103/PhysRevB.54.17759} {\bibfield  {journal} {\bibinfo
  {journal} {Phys. Rev. B}\ }\textbf {\bibinfo {volume} {54}},\ \bibinfo
  {pages} {17759} (\bibinfo {year} {1996})}\BibitemShut {NoStop}%
\bibitem [{\citenamefont {Beyer}(2003)}]{Beyer2003}%
  \BibitemOpen
  \bibfield  {author} {\bibinfo {author} {\bibfnamefont {W.}~\bibnamefont
  {Beyer}},\ }\bibfield  {title} {\bibinfo {title} {Diffusion and evolution of
  hydrogen in hydrogenated amorphous and microcrystalline silicon},\ }\href
  {https://doi.org/10.1016/S0927-0248(02)00438-5} {\bibfield  {journal}
  {\bibinfo  {journal} {Solar Energy Materials and Solar Cells}\ }\textbf
  {\bibinfo {volume} {78}},\ \bibinfo {pages} {235 } (\bibinfo {year}
  {2003})}\BibitemShut {NoStop}%
\bibitem [{\citenamefont {Laaziri}\ \emph
  {et~al.}(1999{\natexlab{a}})\citenamefont {Laaziri}, \citenamefont {Kycia},
  \citenamefont {Roorda}, \citenamefont {Chicoine}, \citenamefont {Robertson},
  \citenamefont {Wang},\ and\ \citenamefont {Moss}}]{laaz}%
  \BibitemOpen
  \bibfield  {author} {\bibinfo {author} {\bibfnamefont {K.}~\bibnamefont
  {Laaziri}}, \bibinfo {author} {\bibfnamefont {S.}~\bibnamefont {Kycia}},
  \bibinfo {author} {\bibfnamefont {S.}~\bibnamefont {Roorda}}, \bibinfo
  {author} {\bibfnamefont {M.}~\bibnamefont {Chicoine}}, \bibinfo {author}
  {\bibfnamefont {J.~L.}\ \bibnamefont {Robertson}}, \bibinfo {author}
  {\bibfnamefont {J.}~\bibnamefont {Wang}},\ and\ \bibinfo {author}
  {\bibfnamefont {S.~C.}\ \bibnamefont {Moss}},\ }\bibfield  {title} {\bibinfo
  {title} {High-energy x-ray diffraction study of pure amorphous silicon},\
  }\href {https://doi.org/10.1103/PhysRevB.60.13520} {\bibfield  {journal}
  {\bibinfo  {journal} {Phys. Rev. B}\ }\textbf {\bibinfo {volume} {60}},\
  \bibinfo {pages} {13520} (\bibinfo {year} {1999}{\natexlab{a}})}\BibitemShut
  {NoStop}%
\bibitem [{not()}]{note1}%
  \BibitemOpen
  \bibinfo {note} {In the literature of {\asi}, one occasionally compares the
  experimental value of the average coordination number of 3.88 from
  Ref.~\onlinecite{laaz} with the calculated value obtained from integrating
  the first peak of the PCF of computer-generated models. However, for
  finite-size models, such a comparison is ill-advised as it leads to a
  considerable presence of {\em isolated} dangling bonds in amorphous networks,
  which results in a pseudo-gap or gapless electronic spectrum.}\BibitemShut
  {Stop}%
\bibitem [{\citenamefont {Laaziri}\ \emph
  {et~al.}(1999{\natexlab{b}})\citenamefont {Laaziri}, \citenamefont {Kycia},
  \citenamefont {Roorda}, \citenamefont {Chicoine}, \citenamefont {Robertson},
  \citenamefont {Wang},\ and\ \citenamefont {Moss}}]{Laaziri}%
  \BibitemOpen
  \bibfield  {author} {\bibinfo {author} {\bibfnamefont {K.}~\bibnamefont
  {Laaziri}}, \bibinfo {author} {\bibfnamefont {S.}~\bibnamefont {Kycia}},
  \bibinfo {author} {\bibfnamefont {S.}~\bibnamefont {Roorda}}, \bibinfo
  {author} {\bibfnamefont {M.}~\bibnamefont {Chicoine}}, \bibinfo {author}
  {\bibfnamefont {J.~L.}\ \bibnamefont {Robertson}}, \bibinfo {author}
  {\bibfnamefont {J.}~\bibnamefont {Wang}},\ and\ \bibinfo {author}
  {\bibfnamefont {S.~C.}\ \bibnamefont {Moss}},\ }\bibfield  {title} {\bibinfo
  {title} {High resolution radial distribution function of pure amorphous
  silicon},\ }\href {https://doi.org/10.1103/PhysRevLett.82.3460} {\bibfield
  {journal} {\bibinfo  {journal} {Phys. Rev. Lett.}\ }\textbf {\bibinfo
  {volume} {82}},\ \bibinfo {pages} {3460} (\bibinfo {year}
  {1999}{\natexlab{b}})}\BibitemShut {NoStop}%
\bibitem [{\citenamefont {Street}(1991)}]{street}%
  \BibitemOpen
  \bibfield  {author} {\bibinfo {author} {\bibfnamefont {R.~A.}\ \bibnamefont
  {Street}},\ }\href@noop {} {\emph {\bibinfo {title} {Hydrogenated Amorphous
  Silicon}}}\ (\bibinfo  {publisher} {Cambridge University Press},\ \bibinfo
  {year} {1991})\BibitemShut {NoStop}%
\bibitem [{\citenamefont {Dahal}\ \emph {et~al.}(2022)\citenamefont {Dahal},
  \citenamefont {Elliott},\ and\ \citenamefont {Biswas}}]{Dahal}%
  \BibitemOpen
  \bibfield  {author} {\bibinfo {author} {\bibfnamefont {D.}~\bibnamefont
  {Dahal}}, \bibinfo {author} {\bibfnamefont {S.~R.}\ \bibnamefont {Elliott}},\
  and\ \bibinfo {author} {\bibfnamefont {P.}~\bibnamefont {Biswas}},\
  }\bibfield  {title} {\bibinfo {title} {Extended-range order in tetrahedral
  amorphous semiconductors: The case of amorphous silicon},\ }\href
  {https://doi.org/10.1103/PhysRevB.105.115203} {\bibfield  {journal} {\bibinfo
   {journal} {Phys. Rev. B}\ }\textbf {\bibinfo {volume} {105}},\ \bibinfo
  {pages} {115203} (\bibinfo {year} {2022})}\BibitemShut {NoStop}%
\bibitem [{\citenamefont {Wright}\ \emph {et~al.}(2007)\citenamefont {Wright},
  \citenamefont {Hannon}, \citenamefont {Sinclair}, \citenamefont {Brunier},
  \citenamefont {Guy}, \citenamefont {Stewart}, \citenamefont {Strobel},\ and\
  \citenamefont {Jansen}}]{Wright}%
  \BibitemOpen
  \bibfield  {author} {\bibinfo {author} {\bibfnamefont {A.~C.}\ \bibnamefont
  {Wright}}, \bibinfo {author} {\bibfnamefont {A.~C.}\ \bibnamefont {Hannon}},
  \bibinfo {author} {\bibfnamefont {R.~N.}\ \bibnamefont {Sinclair}}, \bibinfo
  {author} {\bibfnamefont {T.~M.}\ \bibnamefont {Brunier}}, \bibinfo {author}
  {\bibfnamefont {C.~A.}\ \bibnamefont {Guy}}, \bibinfo {author} {\bibfnamefont
  {R.~J.}\ \bibnamefont {Stewart}}, \bibinfo {author} {\bibfnamefont {M.~B.}\
  \bibnamefont {Strobel}},\ and\ \bibinfo {author} {\bibfnamefont
  {F.}~\bibnamefont {Jansen}},\ }\bibfield  {title} {\bibinfo {title} {Neutron
  scattering studies of hydrogenated, deuterated and fluorinated amorphous
  silicon},\ }\href {http://stacks.iop.org/0953-8984/19/i=41/a=415109}
  {\bibfield  {journal} {\bibinfo  {journal} {J. Phys.: Cond. Matt.}\ }\textbf
  {\bibinfo {volume} {19}},\ \bibinfo {pages} {415109} (\bibinfo {year}
  {2007})}\BibitemShut {NoStop}%
\bibitem [{tgr()}]{tgr}%
  \BibitemOpen
  \bibinfo {note} {For a binary system, the neutron-weighted total PCF, $g(r)$,
  is given by $g(r) = \sum_{ij} \omega_{ij} g_{ij}$, where $\omega_{ij}$ are
  the scale factors involving concentrations and neutron scattering factors of
  the constituent atoms of the system. See Ref.~\onlinecite{SREbook} for
  details.}\BibitemShut {Stop}%
\bibitem [{\citenamefont {Beeman}\ \emph {et~al.}(1985)\citenamefont {Beeman},
  \citenamefont {Tsu},\ and\ \citenamefont {Thorpe}}]{Beeman:1985}%
  \BibitemOpen
  \bibfield  {author} {\bibinfo {author} {\bibfnamefont {D.}~\bibnamefont
  {Beeman}}, \bibinfo {author} {\bibfnamefont {R.}~\bibnamefont {Tsu}},\ and\
  \bibinfo {author} {\bibfnamefont {M.~F.}\ \bibnamefont {Thorpe}},\ }\bibfield
   {title} {\bibinfo {title} {Structural information from the raman spectrum of
  amorphous silicon},\ }\href {https://doi.org/10.1103/PhysRevB.32.874}
  {\bibfield  {journal} {\bibinfo  {journal} {Phys. Rev. B}\ }\textbf {\bibinfo
  {volume} {32}},\ \bibinfo {pages} {874} (\bibinfo {year} {1985})}\BibitemShut
  {NoStop}%
\bibitem [{vno()}]{vnote}%
  \BibitemOpen
  \bibinfo {note} {Care must be taken to calculate the Voronoi volume
  associated with an undercoordinated Si atom that subtends a very narrow/wide
  bond angle with its given neighbors. In some cases, the Voronoi volume of
  such a site is ill-defined and a suitable alternative definition is needed to
  obtain an estimate of the volume.}\BibitemShut {Stop}%
\bibitem [{\citenamefont {Sekimoto}\ \emph {et~al.}(2016)\citenamefont
  {Sekimoto}, \citenamefont {Matsumoto}, \citenamefont {Sagara}, \citenamefont
  {Hishida},\ and\ \citenamefont {Terakawa}}]{Sekimoto2016}%
  \BibitemOpen
  \bibfield  {author} {\bibinfo {author} {\bibfnamefont {T.}~\bibnamefont
  {Sekimoto}}, \bibinfo {author} {\bibfnamefont {M.}~\bibnamefont {Matsumoto}},
  \bibinfo {author} {\bibfnamefont {A.}~\bibnamefont {Sagara}}, \bibinfo
  {author} {\bibfnamefont {M.}~\bibnamefont {Hishida}},\ and\ \bibinfo {author}
  {\bibfnamefont {A.}~\bibnamefont {Terakawa}},\ }\bibfield  {title} {\bibinfo
  {title} {Changes in the vacancy size distribution induced by non-bonded
  hydrogens in hydrogenated amorphous silicon},\ }\href
  {https://doi.org/https://doi.org/10.1016/j.jnoncrysol.2016.05.030} {\bibfield
   {journal} {\bibinfo  {journal} {J. Non-Cryst. Solids}\ }\textbf {\bibinfo
  {volume} {447}},\ \bibinfo {pages} {207 } (\bibinfo {year}
  {2016})}\BibitemShut {NoStop}%
\bibitem [{spa()}]{sparse}%
  \BibitemOpen
  \bibinfo {note} {The radius and the number of H atoms used here to define a
  hydrogen cluster are somewhat arbitrary. One typically chooses a value of
  4--5 {\AA} for radius and 5--7 H atoms to define a cluster. Although the
  clusters are not unique, the total number of H atoms in a clustered
  environment remains more or less the same for a given radius and a minimum
  cluster size.}\BibitemShut {Stop}%
\bibitem [{\citenamefont {Van~Vleck}(1948)}]{Vleck}%
  \BibitemOpen
  \bibfield  {author} {\bibinfo {author} {\bibfnamefont {J.~H.}\ \bibnamefont
  {Van~Vleck}},\ }\bibfield  {title} {\bibinfo {title} {The dipolar broadening
  of magnetic resonance lines in crystals},\ }\href
  {https://doi.org/10.1103/PhysRev.74.1168} {\bibfield  {journal} {\bibinfo
  {journal} {Phys. Rev.}\ }\textbf {\bibinfo {volume} {74}},\ \bibinfo {pages}
  {1168} (\bibinfo {year} {1948})}\BibitemShut {NoStop}%
\bibitem [{\citenamefont {Abragam}(1994)}]{Abragambook}%
  \BibitemOpen
  \bibfield  {author} {\bibinfo {author} {\bibfnamefont {A.}~\bibnamefont
  {Abragam}},\ }\href@noop {} {\emph {\bibinfo {title} {Principles of Nuclear
  Magnetism}}}\ (\bibinfo  {publisher} {Oxford Science Publications},\ \bibinfo
  {year} {1994})\BibitemShut {NoStop}%
\bibitem [{\citenamefont {Timilsina}\ and\ \citenamefont
  {Biswas}(2013)}]{biswas-nmr}%
  \BibitemOpen
  \bibfield  {author} {\bibinfo {author} {\bibfnamefont {R.}~\bibnamefont
  {Timilsina}}\ and\ \bibinfo {author} {\bibfnamefont {P.}~\bibnamefont
  {Biswas}},\ }\bibfield  {title} {\bibinfo {title} {A study of hydrogen
  microstructure in amorphous silicon via inversion of nuclear magnetic
  resonance spectra},\ }\href
  {http://stacks.iop.org/0953-8984/25/i=16/a=165801} {\bibfield  {journal}
  {\bibinfo  {journal} {J. Phys.: Cond. Matt.}\ }\textbf {\bibinfo {volume}
  {25}},\ \bibinfo {pages} {165801} (\bibinfo {year} {2013})}\BibitemShut
  {NoStop}%
\bibitem [{\citenamefont {Tauc}\ \emph {et~al.}(1966)\citenamefont {Tauc},
  \citenamefont {Grigorovici},\ and\ \citenamefont {Vancu}}]{Tauc:1966}%
  \BibitemOpen
  \bibfield  {author} {\bibinfo {author} {\bibfnamefont {J.}~\bibnamefont
  {Tauc}}, \bibinfo {author} {\bibfnamefont {R.}~\bibnamefont {Grigorovici}},\
  and\ \bibinfo {author} {\bibfnamefont {A.}~\bibnamefont {Vancu}},\ }\bibfield
   {title} {\bibinfo {title} {Optical properties and electronic structure of
  amorphous germanium},\ }\href {https://doi.org/10.1002/pssb.19660150224}
  {\bibfield  {journal} {\bibinfo  {journal} {Physica Status Solidi B}\
  }\textbf {\bibinfo {volume} {15}},\ \bibinfo {pages} {627} (\bibinfo {year}
  {1966})}\BibitemShut {NoStop}%
\bibitem [{\citenamefont {Tauc}(1968)}]{Tauc:1968}%
  \BibitemOpen
  \bibfield  {author} {\bibinfo {author} {\bibfnamefont {J.}~\bibnamefont
  {Tauc}},\ }\bibfield  {title} {\bibinfo {title} {Optical properties and
  electronic structure of amorphous {Ge} and {Si}},\ }\href
  {https://doi.org/https://doi.org/10.1016/0025-5408(68)90023-8} {\bibfield
  {journal} {\bibinfo  {journal} {Materials Research Bulletin}\ }\textbf
  {\bibinfo {volume} {3}},\ \bibinfo {pages} {37} (\bibinfo {year}
  {1968})}\BibitemShut {NoStop}%
\bibitem [{\citenamefont {Morigaki}(1991)}]{morigaki}%
  \BibitemOpen
  \bibfield  {author} {\bibinfo {author} {\bibfnamefont {K.}~\bibnamefont
  {Morigaki}},\ }\href@noop {} {\emph {\bibinfo {title} {Physics of Amorphous
  Semiconductors}}}\ (\bibinfo  {publisher} {Imperial College Press},\ \bibinfo
  {year} {1991})\BibitemShut {NoStop}%
\bibitem [{\citenamefont {Elliott}(1988)}]{SREbook}%
  \BibitemOpen
  \bibfield  {author} {\bibinfo {author} {\bibfnamefont {S.~R.}\ \bibnamefont
  {Elliott}},\ }\href@noop {} {\emph {\bibinfo {title} {Physics of Amorphous
  Materials}}}\ (\bibinfo  {publisher} {Longman Higher Education},\ \bibinfo
  {year} {1988})\BibitemShut {NoStop}%
\end{thebibliography}
%

\end {document}